\documentclass[aps,showpacs,showkeys,twocolumn]{revtex4}%
\pdfoutput=1
\usepackage{graphics,amsmath,amsfonts,amscd,revsymb,latexsym,
enumerate,multirow,epsfig}
\usepackage{amsmath}
\usepackage{amsfonts}
\usepackage{amssymb}
\usepackage{graphicx}%
\providecommand{\U}[1]{\protect\rule{.1in}{.1in}}
\providecommand{\U}[1]{\protect\rule{.1in}{.1in}}
\newtheorem{theorem}{Theorem}

\newtheorem{definition}{Definition}

\newtheorem{proposition}[theorem]{Proposition}

\newcommand{\qed}{{\hfill$\Box$}}
\newenvironment{proof}{\noindent \textbf{{Proof~} }}{\qed}
\def\bi{\begin{itemize}}
\def\ei{\end{itemize}}
\def\be{\begin{equation}}
\def\ee{\end{equation}}
\def\bea{\begin{eqnarray}}
\def\eea{\end{eqnarray}}
\def\ben{\begin{eqnarray*}}
\def\een{\end{eqnarray*}}

\def\>{\rangle}
\def\<{\langle}

\newcommand{\1} I

 \DeclareMathOperator{\tr}{Tr}

\def\*{\star}

\def\0{{\mathbf{0}}}
\def\1{{\mathbf{1}}}
\def\2{{\mathbf{2}}}
\def\3{{\mathbf{3}}}
\def\4{{\mathbf{4}}}
\def\5{{\mathbf{5}}}
\def\6{{\mathbf{6}}}
\def\7{{\mathbf{7}}}
\def\8{{\mathbf{8}}}
\def\9{{\mathbf{9}}}

\begin{document}
\title{Public and private communication with a quantum channel and a secret key}
\author{Min-Hsiu Hsieh}
\email{minhsiuh@gmail.com}
\affiliation{ERATO-SORST Quantum Computation and Information Project, Japan Science and
Technology Agency 5-28-3, Hongo, Bunkyo-ku, Tokyo, Japan}
\author{Mark M. Wilde}
\email{mark.m.wilde@saic.com}
\affiliation{Electronic Systems Division, Science Applications International Corporation,
4001 North Fairfax Drive, Arlington, Virginia, USA\ 22203}
\date{\today }


\begin{abstract}
We consider using a secret key and a noisy quantum channel to generate
noiseless public communication and noiseless private communication. The
optimal protocol for this setting is the \textit{publicly-enhanced private
father protocol}. This protocol exploits random coding techniques and
\textquotedblleft piggybacking\textquotedblright\ of public information along
with secret-key-assisted private codes. The publicly-enhanced private father
protocol is a generalization of the secret-key-assisted protocol of Hsieh,
Luo, and Brun and a generelization of a protocol for simultaneous
communication of public and private information suggested by Devetak and Shor.

\end{abstract}
\maketitle

\section{Introduction}

The qualitative connection between secrecy of information and the ability to
maintain quantum correlations has long been a part of quantum information
theory. The connection comes about from the observation that a maximally
entangled ebit state, shared between two parties named Alice and Bob, has no
correlations with the \textquotedblleft rest of the universe\textquotedblright%
---in this sense, the ebit is \textit{monogamous} \cite{T04}. We can represent
the global state of the ebit and the \textquotedblleft rest of the
universe\textquotedblright\ as%
\[
\Phi^{AB}\otimes\sigma^{E},
\]
where Alice and Bob share the ebit $\Phi^{AB}$, and%
\begin{align*}
\Phi^{AB} &  \equiv\left\vert \Phi\right\rangle \left\langle \Phi\right\vert
^{AB},\\
\left\vert \Phi\right\rangle ^{AB} &  \equiv\frac{1}{\sqrt{2}}(\left\vert
0\right\rangle ^{A}\left\vert 0\right\rangle ^{B}+\left\vert 1\right\rangle
^{A}\left\vert 1\right\rangle ^{B}),
\end{align*}
and $\sigma^{E}$ is some state of Eve, a third party representing the
\textquotedblleft rest of the universe.\textquotedblright\ Eve's state
$\sigma^{E}$ is independent of Alice and Bob's ebit. The relation to a secret
key comes about when Alice and Bob perform local measurements of the ebit in
the computational basis. The resulting state is%
\[
\overline{\Phi}^{AB}\otimes\sigma^{E},
\]
where $\overline{\Phi}^{AB}$ is the maximally correlated state:%
\[
\overline{\Phi}^{AB}\equiv\frac{1}{2}\left(  \left\vert 0\right\rangle
\left\langle 0\right\vert ^{A}\otimes\left\vert 0\right\rangle \left\langle
0\right\vert ^{B}+\left\vert 1\right\rangle \left\langle 1\right\vert
^{A}\otimes\left\vert 1\right\rangle \left\langle 1\right\vert ^{B}\right)  .
\]
In this setting, the cryptographic setting, we consider Eve as a potential
eavesdropper. She is no longer the \textquotedblleft rest of the
universe,\textquotedblright\ because some party now holds the purification of
the dephased state $\overline{\Phi}^{AB}$.

The body of literature on the privacy/quantum-coherence connection has now
grown substantially. Some of the original exploitations of this connection
were the various quantum key distribution protocols \cite{BB84,E91,BBM92}.
These protocols establish a shared secret key with the help of a noisy quantum
channel or noisy entanglement. The subsequent proofs \cite{SP00,LD07QKD} for
the security of these protocols rely on the formal mathematical equivalence
between entanglement distillation \cite{BDSW96} and key distillation.
Schumacher and Westmoreland explored the connection with an
information-theoretical study \cite{SW98}---they established a simple relation
between the capacity of a quantum channel for transmitting quantum information
and its utility for quantum key distribution. Collins and Popescu \cite{CP02}
and Gisin \textit{et al}. \cite{GRW02} initiated the formal study of the
connections between entanglement and secret key. Since then, researchers have
determined a method for mapping an entangled state to a probability
distribution with secret correlations \cite{AG05} and have continued to extend
existing quantum results \cite{CVDC03}\ to analogous results for privacy
\cite{bae:032304}.

The connection has also proven fruitful for quantum Shannon theory, where we
study the capabilities of a large number of independent uses of a noisy
quantum channel or a large number of copies of a noisy bipartite state. The
first step in this direction was determining the capacity of a quantum channel
for transmitting a private message or establishing a shared secret key
\cite{Devetak03,CWY04}. Devetak further showed how \textit{coherently}
performing each step of a private protocol leads to a code that achieves the
capacity of a quantum channel for transmitting quantum information
\cite{Devetak03}. Since these initial insights, we have seen how the seemingly
different tasks of distilling secret key, distilling entanglement,
transmitting private information, and transmitting quantum information all
have connections \cite{DW03b}. Oppenheim \textit{et al}. have determined a
merging protocol for private correlations \cite{OSW05}, based on the quantum
state merging protocol \cite{nature2005horodecki,cmp2007HOW}. Additionally,
the secret-key-assisted private capacity of a quantum channel \cite{HLB08SKP}
is analogous to its entanglement-assisted quantum capacity
\cite{DHW03,DHW05RI}.

The connection is only qualitative because the Horodeckis and Oppenheim have
observed that there exist \textit{bound entangled }states \cite{H3O05}. These
bound entangled states are entangled, yet have no distillable entanglement
(one cannot extract ebits from them), but they indeed have distillable secret
key. The dynamic equivalent of this state is an entanglement binding channel
\cite{H300,horodecki:110502,H3LO08}. This channel has no ability to transmit
quantum information. The loss of the privacy-coherence connection here is not
necessarily discomforting. In fact, it is more interesting because it leads to
the \textquotedblleft superactivation effect\textquotedblright%
\ \cite{science2008smith}---the possibility of combining two zero-capacity
channels to form a quantum channel with non-zero quantum capacity.
Additionally, the private analog of this scenario exhibits some unexpected
behavior \cite{smith:010501}.

In this paper, we continue along the privacy-coherence connection and detail
the publicly-enhanced private father protocol. This protocol exploits a secret
key and a large number of independent uses of a noisy quantum channel to
generate noiseless public communication and noiseless private communication.
This protocol is the \textquotedblleft public-private\textquotedblright%
\ analog of the classically-enhanced father protocol \cite{HW08GFP}, and might
lead to further insights into the privacy-coherence connection. The
publicly-enhanced private father protocol combines the coding techniques of
the suggested protocol in Section~4 of Ref.~\cite{DS03}\ (originally proven
for the classical wiretap channel \cite{CK67}) with the recent
secret-key-assisted private communication protocol \cite{HLB08SKP}.

We structure this work as follows. The next section establishes the definition
of a noiseless public channel, a noiseless private channel, noiseless common
randomness, and a perfect secret key. We then clarify a small point with the
protocol for private communication \cite{Devetak03,CWY04}---specifically, we
address the apparent ability of that protocol to transmit public information
in addition to private information. Section~\ref{sec:main-theorem}\ describes
the publicly-enhanced private father protocol and states our main theorem
(Theorem~\ref{thm:PEPFP}). This theorem gives the capacity region for the
publicly-enhanced private father protocol. We proceed with the proof of the
corresponding converse theorem in Section~\ref{sec:converse-theorem}\ and the
proof of the corresponding direct coding theorem in
Section~\ref{sec:direct-coding-theorem}. Section~\ref{sec:children}\ shows
that the suggested protocol from Ref.~\cite{DS03}\ is a child of the
publicly-enhanced private father protocol. We then conclude with some
remaining open questions.

\section{Definitions and Notation}

We first introduce the notion of a noiseless public channel, a noiseless
private channel, and a noiseless secret key as resources. Our communication
model includes one sender Alice, a receiver Bob, and an eavesdropper Eve.
Alice chooses classical messages $k$ from a set $\left[  K\right]
\equiv\left\{  1,\ldots,K\right\}  $. She encodes these messages as quantum
states $\{\left\vert k\right\rangle \left\langle k\right\vert ^{A}%
\}_{k\in\left[  K\right]  }$. We assume that each party is in a local, secret
facility that does not leak information to the outside world. For example, Eve
cannot gain any information about a state that Alice or Bob prepares locally.
We consider two dynamic resources, public classical communicaton and private
classical communication, and two static resources, common randomness and
secret key.

A noiseless public channel id$_{\text{pub}}^{A\rightarrow B}$ from Alice to
Bob implements the following map for $k\in\left[  K\right]  $:%
\begin{equation}
\text{id}_{\text{pub}}^{A\rightarrow B}:\left\vert k\right\rangle \left\langle
k\right\vert ^{A}\rightarrow\left\vert k\right\rangle \left\langle
k\right\vert ^{B}\otimes\sum_{k^{\prime}\in\left[  K\right]  }p_{K^{\prime}%
|K}\left(  k^{\prime}|k\right)  \rho_{k^{\prime}}^{E},
\label{eq:noiseless-public}%
\end{equation}
where $p_{K^{\prime}|K}\left(  k^{\prime}|k\right)  $ is some conditional
probability distribution and $\rho_{k^{\prime}}^{E}$ is a state on Eve's
system. The above definition of a noiseless public channel captures the idea
that Bob receives the classical information perfectly, but Eve receives only
partial information about Alice's message. Eve has perfect correlation with
Alice's message if and only if her conditional distribution $p_{K^{\prime}%
|K}\left(  k^{\prime}|k\right)  $ is $\delta_{k^{\prime},k}$ and her states
$\rho_{k^{\prime}}^{E}=\left\vert k^{\prime}\right\rangle \left\langle
k^{\prime}\right\vert ^{E}$ for all $k^{\prime}$. We make no distinction
between a noiseless public channel where Eve receives partial information and
one where Eve receives perfect information because we are only concerned with
the rate at which Alice can communicate to Bob---we are not concerned with the
more general scenario of broadcast communication where Eve is an active party
in the communication protocol \cite{YHD2006}. We represent the noiseless
public channel symbolically as the following resource:%
\[
\left[  c\rightarrow c\right]  _{\text{pub}}.
\]

The resource inequality framework \cite{DHW05RI} uses the notation $\left[
c\rightarrow c\right]  $ to represent one noiseless bit of classical
communication. We require a symbol different from $\left[  c\rightarrow
c\right]  $ because that symbol does not distinguish between public and
private communication. For example, the superdense coding protocol
\cite{BW92}\ actually produces two private classical bits, but the notation
$\left[  c\rightarrow c\right]  $ does not indicate this fact.

A noiseless private channel is the following map:%
\[
\text{id}_{\text{priv}}^{A\rightarrow B}:\left\vert k\right\rangle
\left\langle k\right\vert ^{A}\rightarrow\left\vert k\right\rangle
\left\langle k\right\vert ^{B}\otimes\sigma^{E},
\]
where $\sigma^{E}$ is a constant state on Eve's system, independent of what
Bob receives. A private channel appears as a special case of a public channel
where random variable $K^{\prime}$ that represents Eve's knowledge is
independent of random variable $K$. The definition in
(\ref{eq:noiseless-public}) reduces to that of a private channel if we set the
probability distribution in (\ref{eq:noiseless-public}) to $p_{K^{\prime}%
|K}\left(  k^{\prime}\right)  $. But we define a private channel as the case
when $K^{\prime}$ and $K$ are independent. Otherwise, the channel is public.
This difference is the distinguishing feature of a noiseless private channel.
We represent the noiseless private channel symbolically as the following
resource:%
\[
\left[  c\rightarrow c\right]  _{\text{priv}}.
\]
The above definitions of a public classical channel and private classical
channel are inspired by definitions in Refs.~\cite{HLB08SKP,LS08}.

\textit{Common randomness} is the static analog of a noiseless public channel
\cite{AC93CR,AC93II,DW03a}. In fact, Alice can actually use a public channel
to implement common randomness. Alice first prepares a local maximally mixed
state $\pi^{A}$ where%
\[
\pi^{A}\equiv\frac{1}{\left\vert K\right\vert }\sum_{k\in\left[  K\right]
}\left\vert k\right\rangle \left\langle k\right\vert ^{A}.
\]
She makes an exact copy of the random state locally to produce the following
state:%
\begin{equation}
\overline{\Phi}^{AA^{\prime}}\equiv\frac{1}{\left\vert K\right\vert }%
\sum_{k\in\left[  K\right]  }\left\vert k\right\rangle \left\langle
k\right\vert ^{A}\otimes\left\vert k\right\rangle \left\langle k\right\vert
^{A^{\prime}}. \label{eq:copied-state}%
\end{equation}
She sends the $A^{\prime}$ system through the noiseless public channel. The
resulting state represents common randomness shared between Alice and Bob,
about which Eve may have partial information:%
\[
\frac{1}{\left\vert K\right\vert }\sum_{k\in\left[  K\right]  }\left\vert
k\right\rangle \left\langle k\right\vert ^{A}\otimes\left\vert k\right\rangle
\left\langle k\right\vert ^{B}\otimes\sum_{k^{\prime}\in\left[  K\right]
}p_{K^{\prime}|K}\left(  k^{\prime}|k\right)  \rho_{k^{\prime}}^{E}%
\]

A noiseless secret key is the static analog of a noiseless private channel.
Alice again prepares the state $\pi^{A}$ and makes a copy of it to an
$A^{\prime}$ system. She sends the $A^{\prime}$ system through a noiseless
private channel, generating the following resource:%
\[
\frac{1}{K}\sum_{k\in\left[  K\right]  }\left\vert k\right\rangle \left\langle
k\right\vert ^{A}\otimes\left\vert k\right\rangle \left\langle k\right\vert
^{B}\otimes\sigma^{E}=\overline{\Phi}^{AB}\otimes\sigma^{E}.
\]
Alice and Bob share perfect common randomness, but this time, Eve has no
knowledge of this common randomness. This resource is a secret key. A perfect
secret key resource has two requirements \cite{Renner:2005:thesis}:

\begin{enumerate}
\item The key should have a uniform distribution.

\item Eve possesses no correlations with the secret key.
\end{enumerate}

We denote the resource of a shared secret key as follows:%
\[
\left[  cc\right]  _{\text{priv}}.
\]

Note that a noiseless public channel alone cannot implement a noiseless
private channel, and a noiseless private channel alone cannot implement a
noiseless public channel. This relation is different from the corresponding
relation between a noiseless quantum channel and a noiseless classical channel
\cite{HW09T3} because a noiseless quantum channel alone can implement a
noiseless classical channel, but a noiseless classical channel alone cannot
implement a noiseless quantum channel.

\section{Relative Resource in Private Communication}

\label{sec:relative-resource-priv}We would like to clarify one point with the
protocol for private communication \cite{Devetak03,CWY04}\ before proceeding
to our main theorem. By inspecting the proof of the direct coding theorem in
Ref.~\cite{Devetak03}, one might think that Alice could actually transmit
public information at an additional rate of $I\left(  X;E\right)  $. The
following sentence from Ref.~\cite{Devetak03} may lead one to arrive at such a conclusion:

\begin{quote}
\textquotedblleft By construction, Bob can perform a measurement that
correctly identifies the pair $\left(  k,m\right)  $, and hence $k$, with
probability $\geq1-\sqrt[4]{\epsilon}$.\textquotedblright
\end{quote}

But this conclusion is incorrect because the random variable $M$ representing
the \textquotedblleft public\textquotedblright\ message $m$ must have a
uniform distribution. This random variable $M$ serves the purpose of
randomizing Eve's knowledge of the private message $k$ \cite{igor09}. The
protocol would not operate as intended if random variable $M$ had a
distribution other than the uniform distribution. The size of the message set
for the random variable $M$ must be at least $2^{nI\left(  X;E\right)  }$. The
rate $I\left(  X;E\right)  $ of randomization further confirms the role of the
mutual information as the minimum amount of noise needed to destroy one's
correlations with a random variable \cite{GPW05} (see Refs.~\cite{B08,B09} for
further explorations of this idea). It is thus not surprising that the mutual
information $I\left(  X;E\right)  $ arises in the protocol for private
communication because Alice would like to destroy Eve's correlations with her
private message $k$.

The resource inequality \cite{DHW05RI} for the protocol for private
communication is as follows:%
\begin{multline}
\left\langle \mathcal{N}\right\rangle \geq I\left(  X;E\right)  \left[
c\rightarrow c:\pi\right]  _{\text{pub}}+\\
\left(  I\left(  X;B\right)  -I\left(  X;E\right)  \right)  \left[
c\rightarrow c\right]  _{\text{priv}},\label{eq:devetak-priv}%
\end{multline}
where the mutual information quantities are with respect to the following
classical-quantum state:%
\[
\sum_{x\in\mathcal{X}}p_{X}\left(  x\right)  \left\vert x\right\rangle
\left\langle x\right\vert ^{X}\otimes U_{\mathcal{N}}^{A^{\prime}\rightarrow
BE}(\sigma_{x}^{A^{\prime}}),
\]
corresponding to the channel input ensemble $\{p_{X}\left(  x\right)
,\sigma_{x}^{A^{\prime}}\}_{x\in\mathcal{X}}$. The meaning of the resource
inequality is that Alice can transmit $nI\left(  X;E\right)  $ bits of public
information (with the requirement that Alice's random variable has a uniform
distribution)\ and $n\left(  I\left(  X;B\right)  -I\left(  X;E\right)
\right)  $ bits of private information by using a large number $n$ of
independent uses of the noisy quantum channel $\mathcal{N}$. The resource
$\left[  c\rightarrow c:\pi\right]  _{\text{pub}}$ is not an absolute
resource, but is rather a \textit{relative resource }\cite{DHW05RI,D06,A06},
meaning that the protocol only works properly if Alice's public variable has a
uniform distribution, or equivalently, is equal to the maximally mixed state
$\pi$. This public information must be completely random because Alice uses it
to randomize Eve's knowledge of the private message.

The resource inequality in (\ref{eq:devetak-priv}) leads to a simpler way of
implementing the direct coding theorem of the secret-key-assisted private
communication protocol \cite{HLB08SKP}. Suppose that Alice has public
information in a random variable $M$. If she combines this random variable
with a secret key, the resulting random variable has a uniform distribution
because the secret key randomizes the public variable. This variable can then
serve as the input needed to implement the relative resource of public
communication. Alice can transmit an extra $nI\left(  X;E\right)  $ of private
information by combining this public communication with the secret key
resource, essentially implementing a one-time pad protocol \cite{V26,S49}. We
phrase the above argument with the theory of resource inequalities:%
\begin{align*}
&  \left\langle \mathcal{N}\right\rangle +I\left(  X;E\right)  \left[
cc\right]  _{\text{priv}}\\
&  \geq I\left(  X;E\right)  \left[  c\rightarrow c:\pi\right]  _{\text{pub}%
}+I\left(  X;E\right)  \left[  cc\right]  _{\text{priv}}+\\
&  \left(  I\left(  X;B\right)  -I\left(  X;E\right)  \right)  \left[
c\rightarrow c\right]  _{\text{priv}}\\
&  \geq I\left(  X;E\right)  \left[  c\rightarrow c\right]  _{\text{priv}%
}+\left(  I\left(  X;B\right)  -I\left(  X;E\right)  \right)  \left[
c\rightarrow c\right]  _{\text{priv}}\\
&  =I\left(  X;B\right)  \left[  c\rightarrow c\right]  _{\text{priv}}.
\end{align*}
This resource inequality is equivalent to that obtained in
Ref.~\cite{HLB08SKP}.

\section{Public and Private Transmission with a Secret Key}

\label{sec:main-theorem}We begin by defining our publicly-enhanced private
father protocol (PEPFP) for a quantum channel $\mathcal{N}^{A^{\prime
}\rightarrow B}$ from a sender Alice to a receiver Bob. The channel has an
extension to an isometry $U_{\mathcal{N}}^{A^{\prime}\rightarrow BE}$, defined
on a bipartite quantum system $BE$, where Bob has access to system $B$ and Eve
has access to system $E$. Alice's task is to transmit, by some large number
$n$ uses of the channel $\mathcal{N}$, one of $K$ public messages and one of
$M$ private messages to Bob. The goal is for Bob to identify the messages with
high probability and for Eve to receive no information about the private
message. In addition, Alice and Bob have access to a private string (a secret
key), picked uniformly at random from the set $\left[  S\right]  $, before the
protocol begins.

An $(n,R,P,R_{S},\epsilon)$ \textit{secret-key-assisted private channel code}
consists of six steps: preparation, encryption, channel coding, transmission,
channel decoding, and decryption. We detail each of these steps below.

\textbf{Preparation.} Alice prepares a public message $k$ in a register $K$
and private message $m$ in a register $M$. Each of these has a uniform
distribution:%
\begin{align*}
\pi^{K}  &  \equiv\frac{1}{K}\sum_{k=1}^{K}\left\vert k\right\rangle
\left\langle k\right\vert ^{K},\\
\pi^{M}  &  \equiv\frac{1}{M}\sum_{m=1}^{M}\left\vert m\right\rangle
\left\langle m\right\vert ^{M}.
\end{align*}
Alice also shares the maximally correlated secret key state $\overline{\Phi
}^{S_{A}S_{B}}$ with Bob:%
\[
\overline{\Phi}^{S_{A}S_{B}}\equiv\frac{1}{S}\sum_{s=1}^{S}\left\vert
s\right\rangle \left\langle s\right\vert ^{S_{A}}\otimes\left\vert
s\right\rangle \left\langle s\right\vert ^{S_{B}}.
\]
The overall state after preparation is%
\[
\pi^{K}\otimes\pi^{M}\otimes\overline{\Phi}^{S_{A}S_{B}}.
\]

\textbf{Encryption.} Alice exploits an encryption map%
\[
f:\left[  M\right]  \times\left[  S\right]  \rightarrow\left[  M\right]  .
\]
The encryption map $f$ computes an encrypted variable $f(m,s)$ that depends on
the private message $m$ and the secret key $s$. Furthermore, the encryption
map $f$ satisfies the following conditions:

\begin{enumerate}
\item For all $s_{1},s_{2}\in\left[  S\right]  $ where $s_{1}\neq s_{2}$:%
\[
f(m,s_{1})\neq f(m,s_{2}).
\]

\item For all $m_{1},m_{2}\in\left[  M\right]  $ where $m_{1}\neq m_{2}$:%
\[
f(m_{1},s)\neq f(m_{2},s).
\]

\end{enumerate}

The encryption map $f$ corresponds physically to a CPTP map$\ \mathcal{F}%
^{MS_{A}\rightarrow P}$. The state after the encryption map is%
\begin{multline*}
\mathcal{F}^{MS_{A}\rightarrow P}(\pi^{K}\otimes\pi^{M}\otimes\overline{\Phi
}^{S_{A}S_{B}})=\\
\pi^{K}\otimes\frac{1}{MS}\sum_{m,s}\left\vert f\left(  m,s\right)
\right\rangle \left\langle f\left(  m,s\right)  \right\vert ^{P}%
\otimes\left\vert s\right\rangle \left\langle s\right\vert ^{S_{B}}.
\end{multline*}

\textbf{Channel Encoding.} Alice prepares the codeword state $\sigma
_{k,f\left(  m,s\right)  }^{A^{\prime n}}$ based on the public message $k$ and
the encrypted message $f(m,s)$. This encoding corresponds physically to some
CPTP\ map $\mathcal{E}^{KP\rightarrow A^{\prime n}}$. The state after the
encoding map is%
\[
\frac{1}{KMS}\sum_{k,m,s}\sigma_{k,f\left(  m,s\right)  }^{A^{\prime n}%
}\otimes\left\vert s\right\rangle \left\langle s\right\vert ^{S_{B}}.
\]

\textbf{Transmission.} Alice sends the state $\sigma_{k,f\left(  m,s\right)
}^{A^{\prime n}}$ through the channel $U_{\mathcal{N}}^{A^{\prime
n}\rightarrow B^{n}E^{n}}$, generating the state%
\[
\frac{1}{KMS}\sum_{k,m,s}\sigma_{k,f\left(  m,s\right)  }^{B^{n}E^{n}}%
\otimes\left\vert s\right\rangle \left\langle s\right\vert ^{S_{B}},
\]
where%
\[
\sigma_{k,f\left(  m,s\right)  }^{B^{n}E^{n}}\equiv U_{\mathcal{N}}^{A^{\prime
n}\rightarrow B^{n}E^{n}}(\sigma_{k,f\left(  m,s\right)  }^{A^{\prime n}}).
\]

\textbf{Channel Decoding.} Bob receives the above state from the channel and
would like to decode the messages. He exploits a decoding
positive-operator-valued measure (POVM) that acts on his system $B^{n}$. The
elements of this POVM are%
\[
\{\Lambda_{k,f\left(  m,s\right)  }^{B^{n}}\}_{k\in\left[  K\right]  ,f\left(
m,s\right)  \in\left[  M\right]  }.
\]
Bob places the measurement results $k$ and $f\left(  m,s\right)  $ in the
respective registers $\hat{K}$ and $\hat{P}$. The ideal output state after
Bob's decoding operation is%
\[
\sum_{k,m,s}\sigma_{k,f\left(  m,s\right)  }^{B^{n}E^{n}}\otimes\left\vert
s\right\rangle \left\langle s\right\vert ^{S_{B}}\otimes|k\rangle\langle
k|^{\hat{K}}\otimes|f\left(  m,s\right)  \rangle\langle f\left(  m,s\right)
|^{\hat{P}},
\]
where it is understood that the normalization factor is $1/\left(  KMS\right)
$.

\begin{figure}
[ptb]
\begin{center}
\includegraphics[
width=3.3399in
]%
{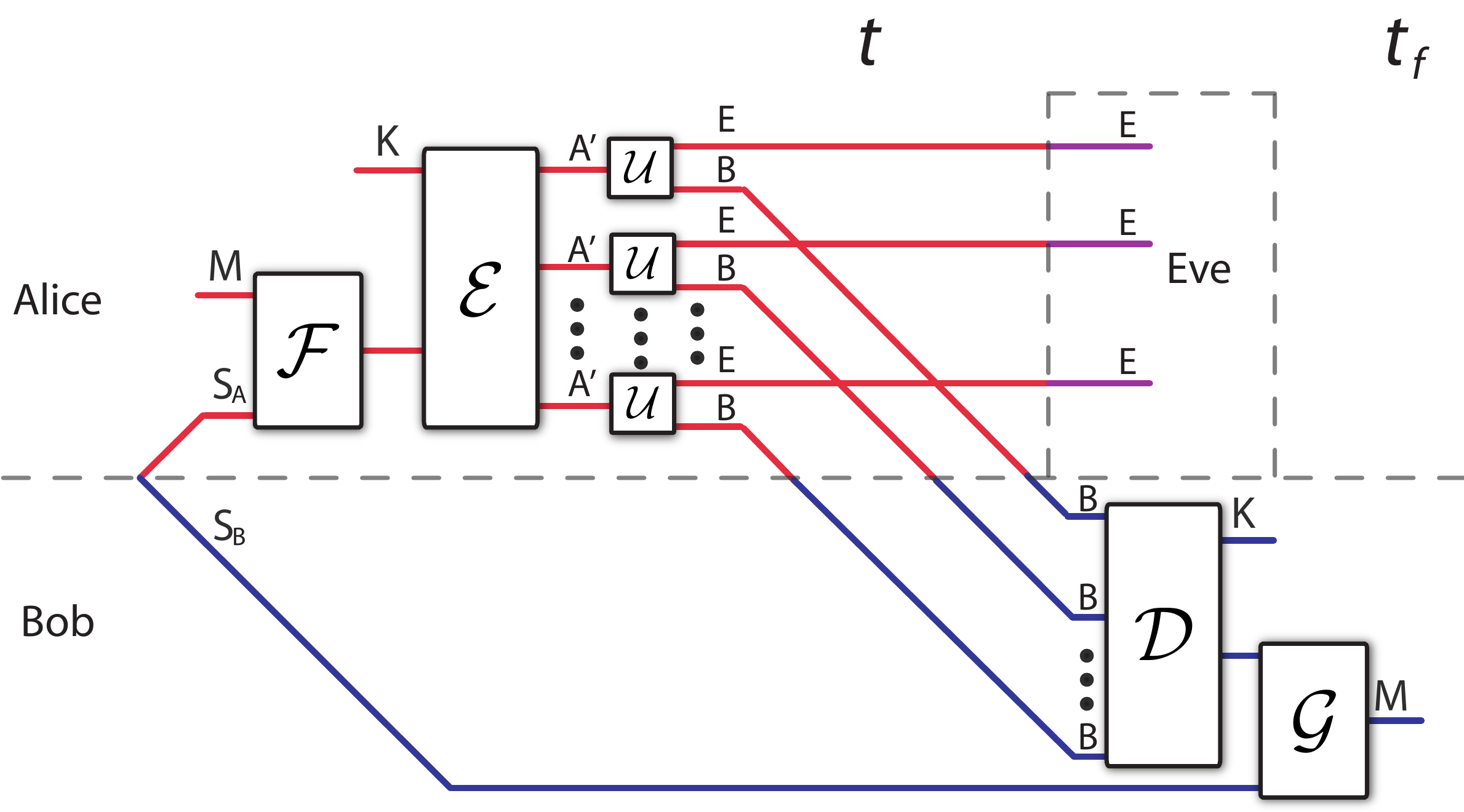}%
\caption{The above figure depicts all of the steps in a publicly-enhanced
private father code. Alice performs the encryption map $\mathcal{F}$ on her
private variable $M$ and her half $S_{A}$ of the secret key. She then encodes
her public variable $K$ and the encrypted message with the encoding map
$\mathcal{E}$. She transmits the encoded data over a large number of uses of
the noisy channel $\mathcal{N}$. The isometric extension of the noisy quantum
channel $\mathcal{N}$ is $U_{\mathcal{N}}$, and we give the full purification
of the channel to Eve. Bob receives the outputs of the channel. He performs
the decoding map $\mathcal{D}$ to recover the public variable $K$ and the
encrypted message. He combines the encrypted message with his half of the
secret key and processes these two variables with the decryption map
$\mathcal{G}$. He then recovers the private variable $M$. A good
publicly-enhanced private father code has the property that Bob can perfectly
recover the public variable $K$ and the private variable $M$ while Eve learns
nothing about the secret key or the private variable $M$.}%
\label{fig:PEPFP}%
\end{center}
\end{figure}

\textbf{Decryption.} The final step is for Bob to decrypt the encrypted
message $f\left(  m,s\right)  $. He employs a decryption function $g$, where%
\[
g:\left[  M\right]  \times\left[  S\right]  \rightarrow\left[  M\right]  .
\]
The decryption function $g$ satisfies the following property:%
\[
\forall~s,m\ \ \ \ \ g(f(m,s),s)=m.
\]
This decryption function allows Bob to recover Alice's private message as
$m=g(f(m,s),s)$ based on the encrypted message $f\left(  m,s\right)  $ and the
secret key $s$. Physically, this operation corresponds to a CPTP\ map
$\mathcal{G}^{S_{B}\hat{P}\rightarrow\hat{M}}$. The state after this
decryption map is%
\[
\frac{1}{KMS}\sum_{k,m,s}\sigma_{k,f\left(  m,s\right)  }^{B^{n}E^{n}}%
\otimes\left\vert s\right\rangle \left\langle s\right\vert ^{S_{B}}%
\otimes|k\rangle\langle k|^{\hat{K}}\otimes|m\rangle\langle m|^{\hat{M}}.
\]
Figure~\ref{fig:PEPFP}\ depicts all of the above steps in a general
publicly-enhanced private father code.%

The conditions for a good publicly-enhanced secret-key-assisted private code
are that Bob be able to decode the public message $k$ and encrypted message
$p=f\left(  m,s\right)  $ with high probability:%
\[
\forall k,p\ \ \ \ \ \text{Tr}\{\Lambda_{k,p}^{B^{n}}\sigma_{k,p}^{B^{n}%
}\}\geq1-\epsilon.
\]
It is sufficient to consider the above criterion because Bob can determine the
private message $m$ with high probability if he can determine the encrypted
message $p$ with high probability. Also, the following inequality is our
security criterion:%
\begin{equation}
\forall k,m\ \ \left\Vert \sum_{s}\sigma_{k,f\left(  m,s\right)  }^{E^{n}%
}\otimes\left\vert s\right\rangle \left\langle s\right\vert ^{S_{B}}%
-\sigma_{k}^{E^{n}}\otimes\pi^{S_{B}}\right\Vert _{1}\leq\epsilon.
\label{eq:security-criterion}%
\end{equation}
This criterion ensures that Eve's state is independent of the key and the
private message $m$.

A rate triple $(R,P,R_{S})$ is \textit{achievable} if there exists an
$(n,R-\delta,P-\delta,R_{S}+\delta,\epsilon)$ publicly-enhanced private father
code for any $\epsilon,\delta>0$ and sufficiently large $n$. The capacity
region $C_{\text{PEPFP}}(\mathcal{N})$ is a three-dimensional region in the
$(R,P,R_{S})$ space with all possible achievable rate triples $(R,P,R_{S})$.

\begin{theorem}
\label{thm:PEPFP} The capacity region $C(\mathcal{\mathcal{N}})$ of a
secret-key-assisted quantum channel $\mathcal{N}$ for simultaneously
transmitting both public and private classical information is equal to the
following expression:%
\begin{equation}
C(\mathcal{N})=\overline{\bigcup_{l=1}^{\infty}\frac{1}{l}C^{(1)}%
(\mathcal{N}^{\otimes l})},\label{pgf}%
\end{equation}
where the overbar indicates the closure of a set. The \textquotedblleft
one-shot\textquotedblright\ region $C^{(1)}(\mathcal{N})$ is the set of all
$R,P,R_{S}\geq0$, such that%
\begin{align}
R &  \leq I(X;B)_{\sigma},\label{pgf1}\\
P &  \leq R_{S}+I\left(  Y;B|X\right)  _{\sigma}-I\left(  Y;E|X\right)
_{\sigma},\label{pgf2}\\
P &  \leq I(Y;B|X)_{\sigma}.\label{pgf3}%
\end{align}
The above entropic quantities are with respect to a \textquotedblleft
one-shot\textquotedblright\ quantum state $\sigma^{XYBE}$, where
\begin{equation}
\sigma^{XYBE}\equiv\sum_{x}p(x)|x\rangle\langle x|^{X}\otimes\rho_{x}%
^{YBE},\label{eq:maximization-state}%
\end{equation}
and the states $\rho_{x}^{YBE}$ are of the form%
\begin{equation}
\rho_{x}^{YBE}=\sum_{y}p(y|x)|y\rangle\langle y|^{Y}\otimes U_{\mathcal{N}%
}^{A^{\prime}\rightarrow BE}(\rho_{x,y}^{A^{\prime}}),\label{eq:state-private}%
\end{equation}
for some density operator $\rho_{x,y}^{A^{\prime}}$ and $U_{\mathcal{N}%
}^{A^{\prime}\rightarrow BE}$ is an isometric extension of $\mathcal{N}$. It
is sufficient to consider $|\mathcal{X}|\leq\min\{|A^{\prime}|,|B|\}^{2}+1$ by
the method in Ref.~\cite{YHD05ieee}.
\end{theorem}

The proof of the above capacity theorem consists of two parts. The first part
that we show is the \textit{converse theorem}. The converse theorem shows that
the rates in the above theorem are optimal---any given coding scheme that has
asymptotically good performance cannot perform any better than the above
rates. We prove the converse theorem in the next section. The second part that
we prove is the \textit{direct coding theorem}. The proof of the direct coding
theorem gives a coding scheme that achieves the limits given in the above theorem.

\section{Proof of the Converse Theorem}

\label{sec:converse-theorem}We outline the proof strategy of the converse
before delving into its details. Consider that a noiseless public channel can
generate common randomness and a noiseless private channel can generate a
secret key. Let $K(\mathcal{N})$ denote the capacity of a quantum channel
$\mathcal{N}$ for generating common randomness, generating a secret key, while
consuming a secret key at respective rates $\left(  R,P,R_{S}\right)  $. The
capacity region $K(\mathcal{N})$ contains the capacity region $C(\mathcal{N})$
of Theorem~\ref{thm:PEPFP}\ ($C(\mathcal{N})\subseteq K(\mathcal{N}%
)$)\ because of the aforementioned one-way relation between a noiseless public
channel and common randomness and that between a noiseless private channel and
a secret key. It thus suffices to prove the converse for a secret-key-assisted
common randomness generation and secret key generation protocol. We consider
the most general such protocol when proving the converse and show that the
capacity region in (\ref{pgf1}-\ref{pgf3}) bounds the capacity region
$K(\mathcal{N})$. The result of the converse theorem is then that
$K(\mathcal{N})\subseteq C(\mathcal{N})$ and thus that $K(\mathcal{N}%
)=C(\mathcal{N})$.%
\begin{figure}
[ptb]
\begin{center}
\includegraphics[
width=3.3399in
]%
{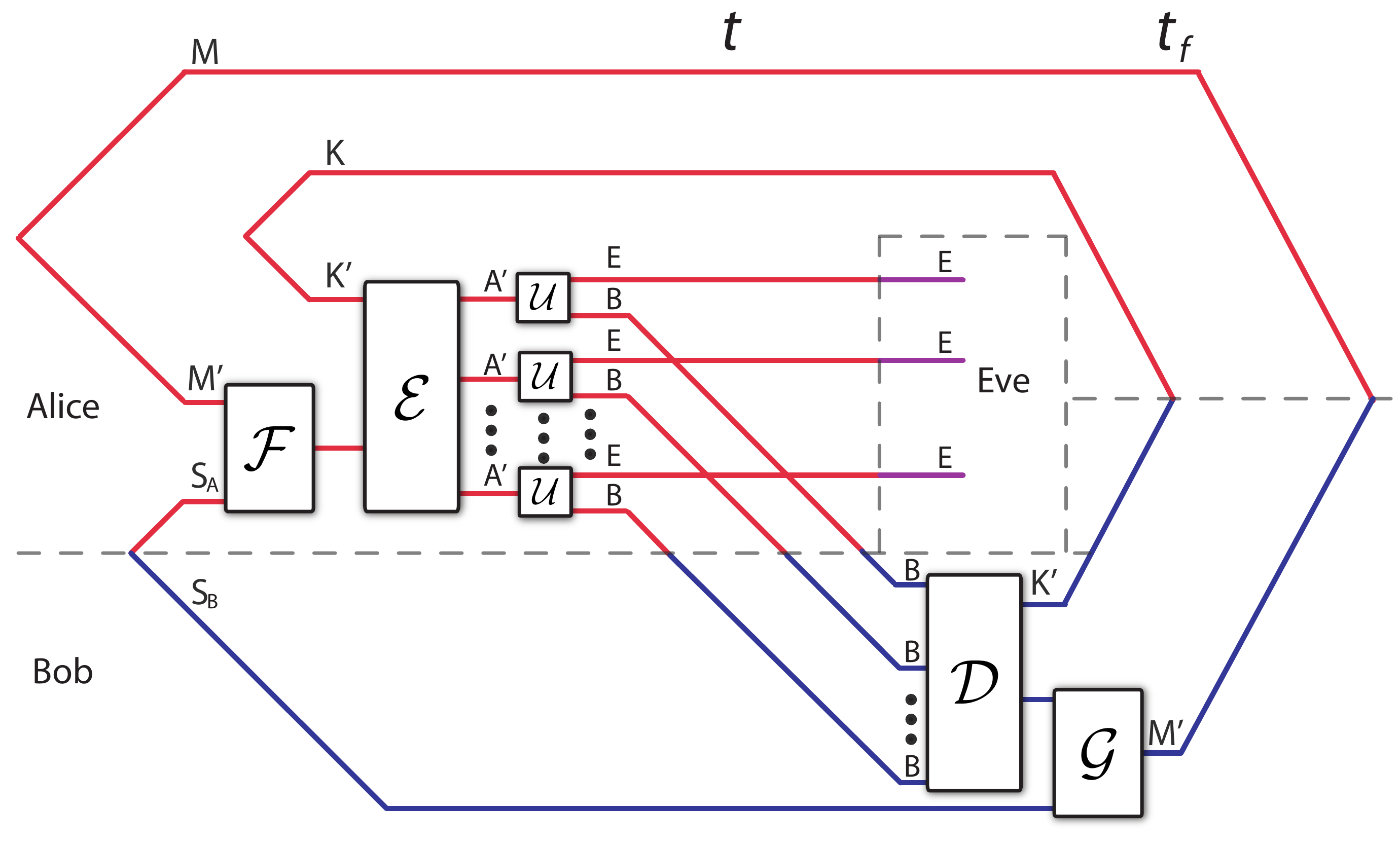}%
\caption{The above figure depicts the coding scenario that we consider for the
converse theorem. It is similar to the protocol of Figure~\ref{fig:PEPFP}~with
the exception that the goal is for Alice and Bob to generate common randomness
and a secret key, rather than transmitting public and private information
respectively.}%
\label{fig:converse}%
\end{center}
\end{figure}

\begin{proof}
[Converse] Suppose Alice creates the maximally correlated state $\pi
^{MM_{A}^{\prime}}$ locally, where%
\[
\overline{\Phi}^{MM_{A}^{\prime}}\equiv\frac{1}{M}\sum_{m=1}^{M}\left\vert
m\right\rangle \left\langle m\right\vert ^{M}\otimes\left\vert m\right\rangle
\left\langle m\right\vert ^{M_{A}^{\prime}}.
\]
(the protocol should be able to transmit the correlations in state
$\overline{\Phi}^{MM_{A}^{\prime}}$ with $\epsilon$-accuracy while keeping
them secret). Alice shares the maximally correlated secret key state
$\overline{\Phi}^{S_{A}S_{B}}$ with Bob:%
\[
\overline{\Phi}^{S_{A}S_{B}}\equiv\frac{1}{S}\sum_{s=1}^{S}\left\vert
s\right\rangle \left\langle s\right\vert ^{S_{A}}\otimes\left\vert
s\right\rangle \left\langle s\right\vert ^{S_{B}}.
\]
Alice prepares a state $\overline{\Phi}^{KK_{A}^{\prime}}$ for common
randomness generation:%
\[
\overline{\Phi}^{KK_{A}^{\prime}}\equiv\frac{1}{K}\sum_{k=1}^{K}\left\vert
k\right\rangle \left\langle k\right\vert ^{K}\otimes\left\vert k\right\rangle
\left\langle k\right\vert ^{K_{A}^{\prime}}.
\]
Alice combines her states $\overline{\Phi}^{KK_{A}^{\prime}}$, $\overline
{\Phi}^{MM_{A}^{\prime}}$, and $\overline{\Phi}^{S_{A}S_{B}}$. The most
general encoding operation that she can perform on her three registers
$K_{A}^{\prime}$, $M_{A}^{\prime}$, and $S_{A}$ is a conditional quantum
encoder $\mathcal{E}^{K_{A}^{\prime}M_{A}^{\prime}S_{A}\rightarrow A^{\prime
n}}$ consisting of a collection $\{\mathcal{E}_{k}^{M_{A}^{\prime}%
S_{A}\rightarrow A^{\prime n}}\}_{k}$\ of CPTP maps \cite{HW08GFP}. Each
element $\mathcal{E}_{k}^{M_{A}^{\prime}S_{A}\rightarrow A^{\prime n}}$ of the
conditional quantum encoder consists of an encryption with the secret key and
the mapping to channel codewords. Each element $\mathcal{E}_{k}^{M_{A}%
^{\prime}S_{A}\rightarrow A^{\prime n}}$ produces the following state:%
\[
\omega_{k}^{MS_{B}A^{\prime n}}\equiv\mathcal{E}_{k}^{M_{A}^{\prime}%
S_{A}\rightarrow A^{\prime n}}(\overline{\Phi}^{MM_{A}^{\prime}}%
\otimes\overline{\Phi}^{S_{A}S_{B}}).
\]
The average density operator over all public messages is then as follows:%
\[
\frac{1}{K}\sum_{k}\left\vert k\right\rangle \left\langle k\right\vert
^{K}\otimes\omega_{k}^{MS_{B}A^{\prime n}}.
\]
Alice sends the $A^{\prime n}$ system through the noisy channel
$U_{\mathcal{N}}^{A^{\prime n}\rightarrow B^{n}E^{n}}$, producing the
following state:%
\[
\omega^{KMS_{B}B^{n}E^{n}}\equiv\frac{1}{K}\sum_{k}\left\vert k\right\rangle
\left\langle k\right\vert ^{K}\otimes U_{\mathcal{N}}^{A^{\prime n}\rightarrow
B^{n}E^{n}}(\omega_{k}^{MS_{B}A^{\prime n}}).
\]
Define the systems $Y\equiv MS_{B}$ and $X\equiv K$ so that the above state is
a particular $n^{\text{th}}$ extension of the state in the statement of the
public-private secret-key-assisted capacity theorem. The above state is the
state at time $t$ in Figure~\ref{fig:converse}. Bob receives the above state
and performs a decoding instrument $\mathcal{D}^{B^{n}S_{B}\rightarrow
K_{B}^{\prime}M_{B}^{\prime}}$ \cite{HW08GFP}\ (each element $\mathcal{D}%
_{k}^{B^{n}S_{B}\rightarrow M_{B}^{\prime}}$ of the instrument consists of a
channel decoding and a decryption). The protocol ends at time $t_{f}%
$\ (depicted in Figure~\ref{fig:converse}). Let $\left(  \omega^{\prime
}\right)  ^{KMK_{B}^{\prime}M_{B}^{\prime}E^{n}}$ be the state at time $t_{f}$
after Bob processes $\omega^{KMS_{B}B^{n}E^{n}}$ with the decoding instrument
$\mathcal{D}^{B^{n}S_{B}\rightarrow K_{B}^{\prime}M_{B}^{\prime}}$.

Suppose that an $\left(  n,R-\delta,P-\delta,R_{S}+\delta,\epsilon\right)  $
secret-key-assisted protocol as given above exists. In particular, the
following information-theoretic security conditions follow from the security
criterion in (\ref{eq:security-criterion}):%
\begin{align}
I\left(  M;E^{n}|K\right)  _{\omega} &  \leq\epsilon
,\label{eq:private-correlations-pub}\\
I\left(  S_{B};E^{n}|K\right)  _{\omega} &  \leq\epsilon
,\label{eq:secret-key-pub}%
\end{align}
by the application of the Alicki-Fannes inequality \cite{0305-4470-37-5-L01}
and evaluating the conditional mutual informations of the ideal state
$\sigma_{k}^{E^{n}}\otimes\pi^{S_{B}}$\ in (\ref{eq:security-criterion}).
These conditions imply that Eve learns nothing about the secret correlations
in system $M$ and Eve learns nothing about the secret key $S_{B}$ (at time
$t$) even if she knows the public variable $K$. We prove that the following
bounds apply to the elements of the protocol's rate triple $\left(
R-\delta,P-\delta,R_{S}+\delta\right)  $,%
\begin{align}
R-\delta &  \leq\frac{I(X;B^{n})_{\omega}}{n},\label{eq:pub-bound}\\
P-\delta &  \leq\frac{I(Y;B^{n}|X)_{\omega}}{n},\label{eq:priv-bound-1}\\
P-\delta &  \leq R_{S}+\frac{I(Y;B^{n}|X)_{\omega}-I(Y;E^{n}|X)_{\omega}}%
{n},\label{eq:priv-bound-2}\\
R_{S}+\delta &  \geq\frac{I(Y;E^{n}|X)_{\omega}}{n},\label{eq:sec-bound}%
\end{align}
for any $\epsilon,\delta>0$ and all sufficiently large $n$.

In the ideal case, the ideal private channel acts on system $M$ to produce the
maximally correlated and secret state $\pi^{MM^{\prime}}$. So, for our case,
the inequality%
\begin{equation}
\left\Vert \left(  \omega^{\prime}\right)  ^{MM_{B}^{\prime}E^{n}}%
-\overline{\Phi}^{MM_{B}^{\prime}}\otimes\sigma^{E^{n}}\right\Vert _{1}%
\leq\epsilon\label{eq:converse-good-private-comm}%
\end{equation}
holds because the protocol is $\epsilon$-good for private communication. The
state $\sigma^{E^{n}}$ is some constant state on Eve's system.

The lower bound in (\ref{eq:sec-bound}) is the most straightforward to prove.
Consider the following chain of inequalities:%
\begin{align*}
&  n\left(  R_{S}+\delta\right)  +2\epsilon\\
&  \geq I\left(  M;E^{n}|K\right)  _{\omega}+I\left(  S_{B};E^{n}|K\right)
_{\omega}+H\left(  S_{B}|K\right)  _{\omega}\\
&  =H\left(  M|K\right)  _{\omega}+H\left(  E^{n}|K\right)  -H\left(
ME^{n}|K\right)  _{\omega}+\\
&  I\left(  S_{B};E^{n}|K\right)  _{\omega}+H\left(  S_{B}|K\right)  _{\omega
}\\
&  \geq H\left(  M|S_{B}K\right)  _{\omega}+H\left(  E^{n}|S_{B}K\right)
-H\left(  ME^{n}|K\right)  _{\omega}+\\
&  I\left(  S_{B};E^{n}|K\right)  _{\omega}+H\left(  S_{B}|K\right)  _{\omega
}\\
&  \geq H\left(  M|S_{B}K\right)  _{\omega}+H\left(  E^{n}|S_{B}K\right)
-H\left(  ME^{n}S_{B}|K\right)  _{\omega}+\\
&  I\left(  S_{B};E^{n}|K\right)  _{\omega}+H\left(  S_{B}|K\right)  _{\omega
}\\
&  =I\left(  M;E^{n}|S_{B}K\right)  _{\omega}+I\left(  S_{B};E^{n}|K\right)
_{\omega}\\
&  =I\left(  MS_{B};E^{n}|K\right)  _{\omega}\\
&  =I\left(  Y;E^{n}|X\right)  _{\omega}%
\end{align*}
The first inequality follows by combining the equality $n\left(  R_{S}%
+\delta\right)  =H\left(  S_{B}\right)  =H\left(  S_{B}|K\right)  $ and the
security criteria in (\ref{eq:private-correlations-pub}%
-\ref{eq:secret-key-pub}). The first equality follows from the definition of
mutual information. The second inequality follows because $H\left(  M\right)
_{\omega}=H\left(  M|S_{B}K\right)  _{\omega}$ ($M$, $S_{B}$, and $K$ are
independent) and conditioning does not increase entropy $H\left(
E^{n}|K\right)  \geq H\left(  E^{n}|S_{B}K\right)  $. The third inequality
follows because the addition of a classical system can increase entropy
$H\left(  ME^{n}|K\right)  _{\omega}\leq H\left(  ME^{n}S_{B}|K\right)
_{\omega}$. The second equality follows from the definition of conditional
mutual information. The third equality follows from the chain rule of mutual
information, and the last equality follows from the definitions $Y\equiv
MS_{B}$ and $X\equiv K$.

We next prove the upper bound in (\ref{eq:priv-bound-1}) on the private
communication rate:%
\begin{align*}
&  n(P-\delta)\\
&  =H\left(  M\right) \\
&  =I\left(  M;M_{B}^{\prime}\right)  _{\omega^{\prime}}+H\left(
M|M_{B}^{\prime}\right) \\
&  \leq I(M;M_{B}^{\prime}K)_{\omega^{\prime}}+n\delta^{\prime}\\
&  \leq I(M;B^{n}S_{B}K)_{\omega}+n\delta^{\prime}\\
&  =I(M;B^{n}K|S_{B})_{\omega}+n\delta^{\prime}\\
&  =H\left(  M|S_{B}\right)  +H(B^{n}K|S_{B})_{\omega}-\\
&  H\left(  MB^{n}S_{B}K\right)  +H\left(  S_{B}\right)  +n\delta^{\prime}\\
&  =H\left(  MS_{B}|K\right)  -H\left(  S_{B}|K\right)  +H(B^{n}%
K|S_{B})_{\omega}-\\
&  H\left(  MB^{n}S_{B}K\right)  +H\left(  S_{B}|K\right)  +n\delta^{\prime}\\
&  =H\left(  MS_{B}|K\right)  +H(B^{n}K|S_{B})_{\omega}-\\
&  H\left(  MB^{n}S_{B}K\right)  +n\delta^{\prime}\\
&  \leq H\left(  MS_{B}|K\right)  +H(B^{n}K)_{\omega}-H\left(  MB^{n}%
S_{B}K\right)  +\\
&  H\left(  K\right)  -H\left(  K\right)  +n\delta^{\prime}\\
&  =I\left(  MS_{B};B^{n}|K\right)  _{\omega}+n\delta^{\prime}\\
&  =I\left(  Y;B^{n}|X\right)  _{\omega}+n\delta^{\prime}%
\end{align*}
The first equality follows by evaluating the entropy for the state
$\overline{\Phi}^{M}$ and noting that $H\left(  M\right)  =H\left(
M|K\right)  $. The second equality follows by standard entropic relations. The
first inequality follows from (\ref{eq:converse-good-private-comm}), Fano's
inequality \cite{CT91}, and conditioning does not increase entropy. The second
inequality is from quantum data processing. The third equality follows from
the chain rule for mutual information and $I\left(  M;S_{B}\right)  =0$
because $M$ and $S_{B}$ are independent. The fourth equality follows by
expanding the conditional mutual information. The fifth and sixth equalities
follow from standard entropic relations. The last inequality follows because
conditioning does not increase entropy $H(B^{n}K|S_{B})_{\omega}\leq
H(B^{n}K)_{\omega}$. The fifth equality follows by the definition of mutual
information, and the last equality follows from the definitions $Y\equiv
MS_{B}$, $X\equiv K$, and $\delta^{\prime}\equiv\frac{1}{n}+\epsilon P$.

The second bound in (\ref{eq:priv-bound-2}) on the private communication rate
follows from adding the bound in (\ref{eq:priv-bound-1}) to the bound in
(\ref{eq:sec-bound}).

We can use a proof by contradiction to get the bound on the public rate $R$.
Suppose that we have secret key available at some rate $>I(X;E^{n})_{\omega
}/n$. Then one could combine the public communication at rate $R$ with the
extra secret key in a one-time pad protocol in order to generate private
communication at a rate $R+P$. The resulting protocol consumes secret key at a
rate greater than $I\left(  YX;E^{n}\right)  $ because%
\[
\frac{I\left(  Y;E^{n}|X\right)  _{\omega}}{n}+\frac{I\left(  X;E^{n}\right)
_{\omega}}{n}=\frac{I\left(  YX;E^{n}\right)  }{n}.
\]
The state $\omega$ is of the form given by the secret-key-assisted capacity
theorem \cite{HLB08SKP}. The total amount of private communication that a
secret-key-assisted protocol can generate cannot be any larger than $I\left(
YX;B^{n}\right)  /n$ \cite{HLB08SKP}. The chain rule also applies to the
mutual information $I\left(  YX;B^{n}\right)  /n$:%
\[
\frac{I\left(  Y;B^{n}|X\right)  _{\omega}}{n}+\frac{I\left(  X;B^{n}\right)
_{\omega}}{n}=\frac{I\left(  YX;B^{n}\right)  }{n}.
\]
If the public rate $R$ were to exceed $I\left(  X;B^{n}\right)  _{\omega}/n$,
then this public rate would contradict the optimality of the
secret-key-assisted protocol from Ref.~\cite{HLB08SKP}. Thus, the public rate
$R$ must obey the bound in (\ref{eq:pub-bound}).
\end{proof}

\section{Proof of the Direct Coding Theorem}

\label{sec:direct-coding-theorem}The direct coding theorem is the proof of the
following \textit{publicly-enhanced private father protocol resource
inequality} (See Refs.~\cite{DHW03,DHW05RI}\ for the theory of resource
inequalities):%
\begin{multline}
\left\langle \mathcal{N}\right\rangle +I\left(  Y;E|X\right)  _{\sigma}\left[
cc\right]  _{\text{priv}}\geq\\
I\left(  Y;B|X\right)  _{\sigma}\left[  c\rightarrow c\right]  _{\text{priv}%
}+I\left(  X;B\right)  _{\sigma}\left[  c\rightarrow c\right]  _{\text{pub}}.
\label{eq:PEPFP-resource-ineq}%
\end{multline}
The resource inequality has an interpretation as the following statement. For
any $\epsilon,\delta>0$ and sufficiently large $n$, there exists a protocol
that consumes $nI\left(  Y;E|X\right)  _{\sigma}$ bits of secret key and $n$
independent uses of the noisy quantum channel $\mathcal{N}$ to generate
$nI\left(  Y;B|X\right)  _{\sigma}$ bits of private communication and
$nI\left(  X;B\right)  _{\sigma}$ bits of public communication with $\epsilon$
probability of error. In addition, Eve's state is $\epsilon$-close to a state
that is independent of the private message and the secret key. The entropic
quantities are with respect to the state $\sigma^{XYBE}$ in
(\ref{eq:maximization-state}).

The proof of the direct coding theorem proceeds similarly to the proof of the
direct coding theorem for the classically-enhanced father protocol from
Ref.~\cite{HW08GFP}. There are some subtle differences between the two proofs,
and we highlight only the parts of the proof that are different from the proof
of the classically-enhanced father protocol. The proof begins by showing how
to construct a \textit{random private father code}, similar to the notion of a
random father code \cite{HW08GFP} or a random quantum code \cite{Devetak03}.
We introduce the \textit{channel input density operator} for a random private
father code and show that it is possible to make it close to a tensor-product
state. We then show how to associate a classical string to a random private
father code by exploiting the \textquotedblleft code pasting\textquotedblright%
\ technique from Ref.~\cite{DS03}. The proof proceeds by applying the
HSW\ theorem \cite{Hol98,SW97}\ to show that Bob can decode the public
information first. Based on the public information, Bob decodes the private
information. The details of the proof involve showing how the random
publicly-enhanced private father code has low probability of error for
decoding the public information and the private information. Finally, we
employ the standard techniques of derandomization and expurgation to show that
there exists a particular publicly-enhanced private father code that achieves
the rates given in Theorem~\ref{thm:PEPFP}.

\subsection{Random Private Coding}

We first recall the secret-key-assisted private communication capacity theorem
(also known as the private father capacity theorem) \cite{HLB08SKP}.

\begin{theorem}
\label{pf_2}The secret-key-assisted private channel capacity region
$C_{\text{SKP}}(\mathcal{N})$ is given by
\begin{equation}
C_{\text{\emph{SKP}}}(\mathcal{N})=\overline{\bigcup_{l=1}^{\infty}\frac{1}%
{l}\widetilde{C}_{\text{\emph{SKP}}}^{(1)}(\mathcal{N}^{\otimes l})},
\label{cp2}%
\end{equation}
where the overbar indicates the closure of a set, and $\widetilde
{C}_{\text{\emph{SKP}}}^{(1)}(\mathcal{N})$ is the set of all $R_{S}\geq0$,
$P\geq0$ such that
\begin{align}
P  &  \leq I(Y;B)_{\rho}-I(Y;E)_{\rho}+R_{S}\label{thm2cond1}\\
P  &  \leq I(Y;B)_{\rho}, \label{thm2cond2}%
\end{align}
where $R_{S}$ is the secret key consumption rate and $\rho$ is a state of the
form
\begin{equation}
\rho^{YBE}\equiv\sum_{y}p(y)|y\rangle\langle y|^{Y}\otimes U_{\mathcal{N}%
}^{A^{\prime}\rightarrow BE}(\rho_{y}^{A^{\prime}}), \label{eq:private-state}%
\end{equation}
for some ensemble $\{p(y),\rho_{y}^{A^{\prime}}\}$ and $U_{\mathcal{N}%
}^{A^{\prime}\rightarrow BE}$ is an isometric extension of $\mathcal{N}$.
\end{theorem}

The \textit{channel input density operator} $\rho^{A^{\prime n}}\left(
\mathcal{C}\right)  $ for a private father code $\mathcal{C}\equiv\{\rho
_{m}^{A^{\prime n}}\}_{m\in\left[  M\right]  }$ is a uniform mixture of all
the private codewords $\rho_{m}^{A^{\prime n}}$\ in code $\mathcal{C}$:%
\[
\rho^{A^{\prime n}}\left(  \mathcal{C}\right)  \equiv\frac{1}{M}\sum_{m=1}%
^{M}\rho_{m}^{A^{\prime n}}.
\]

We cannot say much about the channel input density operator $\rho^{A^{\prime
n}}\left(  \mathcal{C}\right)  $ for a particular private father code
$\mathcal{C}$. But we can say something about the expected channel input
density operator of a \textit{random private father code} $\mathcal{C}$ (where
$\mathcal{C}$ itself becomes a random variable).

\begin{definition}
A \textit{random private father code} is an ensemble $\{p_{\mathcal{C}%
},\mathcal{C}\}$ of codes where each code $\mathcal{C}$ occurs with
probability $p_{\mathcal{C}}$. The \textit{expected channel input density
operator} $\overline{\rho}^{A^{\prime n}}$ is as follows:
\begin{equation}
\overline{\rho}^{A^{\prime n}}\equiv\mathbb{E}_{\mathcal{C}}\left\{
\rho^{A^{\prime n}}\left(  \mathcal{C}\right)  \right\}  .
\label{eq:def_expected_state}%
\end{equation}
A random private father code is \textquotedblleft$\rho$-like\textquotedblright%
\ if the expected channel input density operator is close to a tensor power of
some state $\rho$:%
\begin{equation}
\left\Vert \overline{\rho}^{A^{\prime n}}-\rho^{\otimes n}\right\Vert _{1}%
\leq\epsilon.
\end{equation}

\end{definition}

We now state a version of the direct coding theorem that applies to random
private father codes. The proof shows that we can produce a random
secret-key-assisted private code with an expected channel input density
operator close to a tensor power state.

\begin{proposition}
\label{thm:random-private-code}For any $\epsilon,\delta>0$ and all
sufficiently large $n$, there exists a random $\rho^{A^{\prime}}$-like
secret-key-assisted private code for a channel $\mathcal{N}^{A^{\prime
}\rightarrow B}$ such that%
\begin{equation}
\left\Vert \overline{\rho}^{A^{\prime n}}-(\rho^{A^{\prime}})^{\otimes
n}\right\Vert _{1}\leq2\epsilon+4\sqrt[4]{\epsilon},
\end{equation}
where $\overline{\rho}^{A^{\prime n}}$ is defined in
(\ref{eq:def_expected_state}). The random private code has private
communication rate $I(Y;B)_{\rho}-\delta$ and secret key consumption rate
$I(Y;E)_{\rho}+\delta$. The entropic quantities are with respect to the state
in (\ref{eq:private-state}) and the state $\rho^{A^{\prime}}\equiv\sum
_{y}p\left(  y\right)  \rho_{y}^{A^{\prime}}$.
\end{proposition}

\label{AP_RG0}The proof of Proposition~\ref{thm:random-private-code} is an
extension of the development in Appendix~D of Ref.~\cite{DS03} and the
development in Ref.~\cite{HLB08SKP}.

\begin{proof}
Consider the density operator $\rho^{A^{\prime}}$ where%
\[
\rho^{A^{\prime}}=\sum_{y\in\mathcal{Y}}p\left(  y\right)  \rho_{y}%
^{A^{\prime}}.
\]
The $n^{\text{th}}$ extension of the above state as a tensor power state is as
follows:
\[
\rho^{A^{\prime n}}\equiv(\rho^{A^{\prime}})^{\otimes n}=\sum_{y^{n}%
\in\mathcal{Y}^{n}}p^{n}\left(  y^{n}\right)  \rho_{y^{n}}^{A^{\prime n}},
\]
where%
\[
\rho_{y^{n}}^{A^{\prime n}}\equiv\rho_{y_{1}}^{A^{\prime}}\otimes\rho_{y_{2}%
}^{A^{\prime}}\otimes\cdots\otimes\rho_{y_{n}}^{A^{\prime}}.
\]
We define the pruned distribution $p^{\prime n}$ as follows:%
\[
p^{\prime n}\left(  x^{n}\right)  \equiv\left\{
\begin{array}
[c]{ccc}%
p^{n}\left(  y^{n}\right)  /\sum_{y^{n}\in T_{\delta}^{Y^{n}}}p^{n}\left(
y^{n}\right)  & : & y^{n}\in T_{\delta}^{Y^{n}}\\
0 & : & \text{else},
\end{array}
\right.
\]
where $T_{\delta}^{Y^{n}}$ denotes the $\delta$-typical set of sequences with
length $n$. Let $\widetilde{\rho}^{A^{\prime n}}$ denote the following
\textquotedblleft pruned state\textquotedblright:%
\begin{equation}
\widetilde{\rho}^{A^{\prime n}}\equiv\sum_{y^{n}\in T_{\delta}^{Y^{n}}%
}p^{\prime n}\left(  y^{n}\right)  \rho_{y^{n}}^{A^{\prime n}}.
\label{eq:pruned-state}%
\end{equation}
For any $\epsilon>0$ and sufficiently large $n$, the state $\rho^{A^{\prime
n}}$ is close to $\widetilde{\rho}^{A^{\prime n}}$\ by the gentle measurement
lemma \cite{Winter99} and because the probability for sequences outside the
typical set is small:%
\[
\left\Vert \rho^{A^{\prime n}}-\widetilde{\rho}^{A^{\prime n}}\right\Vert
_{1}\leq2\epsilon.
\]

For any density operator $\rho^{A^{\prime}}$, it is possible to construct a
secret-key-assisted private code that achieves the private communication rate
and secret key consumption rate in Proposition~\ref{thm:random-private-code}.

Let $[M]$ denote a set of size $2^{n[I(Y;B)-c\delta]}$ for some constant $c$
and let $U_{m}$ denote $2^{n[I(Y;B)-c\delta]}$ random variables that we choose
according to the pruned distribution $p^{\prime n}(y^{n})$. The realizations
$u_{m}$ of the random variables $U_{m}$ are sequences in $\mathcal{Y}^{n}$ and
are the basis for constructing a secret-key-assisted private code
$\mathcal{C}$ with the following codeword ensemble:%
\[
\mathcal{C}=\{p^{\prime n}(u_{m}),\rho_{u_{m}}^{A^{n}}\}_{m}.
\]
We then perform a decoding positive operator-valued measure (POVM) with
elements $\{\Lambda_{m}\}_{m\in\left[  M\right]  }$ and decryption map $g$,
resulting in failure with probability $4\epsilon+20\sqrt{\epsilon}$ by the
arguments in Ref.~\cite{HLB08SKP}.

Suppose that we choose a particular secret-key-assisted private code
$\mathcal{C}$ according to the above prescription. Its code density operator
is%
\[
\rho^{A^{\prime n}}(\mathcal{C})=\frac{1}{M}\sum_{m=1}^{M}\rho_{u_{m}%
}^{A^{\prime n}}.
\]

Suppose we now consider the secret-key-assisted private code chosen according
to the above prescription as a \textit{random} code $\mathcal{C}$ (where
$\mathcal{C}$ is now a random variable). Let $\rho^{\prime A^{\prime n}%
}\left(  \mathcal{C}\right)  $ be the channel input density operator for the
random code before expurgation and $\rho^{A^{\prime n}}\left(  \mathcal{C}%
\right)  $ its channel input density operator after expurgation:%
\begin{align*}
\rho^{\prime A^{\prime n}}(\mathcal{C})  &  \equiv\frac{1}{M^{\prime}}%
\sum_{m=1}^{M^{\prime}}\rho_{U_{m}}^{A^{\prime n}},\\
\rho^{A^{\prime n}}(\mathcal{C})  &  \equiv\frac{1}{M}\sum_{m=1}^{M}%
\rho_{U_{m}}^{A^{\prime n}},
\end{align*}
where the primed rates are the rates before expurgation and the unprimed rates
are those after expurgation (they are slightly different but identical for
large $n$). Let $\overline{\rho}^{\prime A^{\prime n}}$ and $\overline{\rho
}^{A^{\prime n}}$ denote the expectation of the above channel input density
operators:%
\begin{align*}
\overline{\rho}^{\prime A^{\prime n}}  &  \equiv\mathbb{E}_{\mathcal{C}%
}\left\{  \rho^{\prime A^{\prime n}}\left(  \mathcal{C}\right)  \right\}  ,\\
\overline{\rho}^{A^{\prime n}}  &  \equiv\mathbb{E}_{\mathcal{C}}\left\{
\rho^{A^{\prime n}}\left(  \mathcal{C}\right)  \right\}  .
\end{align*}

Choosing our code in the particular way that we did leads to an interesting
consequence. The expectation of the density operator corresponding to Alice's
codeword $\rho_{U_{m}}^{A^{\prime n}}$ is equal to the pruned state
in\ (\ref{eq:pruned-state}):
\[
\mathbb{E}_{\mathcal{C}}\left\{  \rho_{U_{m}}^{A^{\prime n}}\right\}
=\sum_{y^{n}}p^{\prime n}(y^{n})\rho_{y^{n}}^{A^{\prime n}},
\]
because we choose the codewords $\rho_{y^{n}}^{A^{\prime n}}$ randomly
according to the pruned distribution $p^{\prime n}(y^{n})$. Then the expected
channel input density operator $\overline{\rho}^{\prime A^{\prime n}}$\ is as
follows:
\begin{align}
\overline{\rho}^{\prime A^{\prime n}} &  =\mathbb{E}_{\mathcal{C}}\left\{
\rho^{\prime A^{\prime n}}\left(  \mathcal{C}\right)  \right\}  \\
&  =\frac{1}{M^{\prime}}\sum_{m=1}^{M^{\prime}}\mathbb{E}_{\mathcal{C}%
}\left\{  \rho_{U_{m}}^{A^{\prime n}}\right\}  \\
&  =\sum_{y^{n}}p^{\prime n}(y^{n})\rho_{y^{n}}^{A^{\prime n}}.
\end{align}
Then we know that the following inequality holds for $\overline{\rho}^{\prime
A^{\prime n}}$ and the tensor power state $\rho^{A^{\prime n}}$%
\begin{equation}
\left\Vert \overline{\rho}^{\prime A^{\prime n}}-\rho^{A^{\prime n}%
}\right\Vert _{1}\leq2\epsilon\label{eq:typical-close}%
\end{equation}
by the typical subspace theorem and the gentle measurement lemma. The
expurgation of any secret-key-assisted private code $\mathcal{C}$ has a
minimal effect on the resulting channel input density operator \cite{DS03}:%
\[
\left\Vert \rho^{\prime A^{\prime n}}\left(  \mathcal{C}\right)
-\rho^{A^{\prime n}}\left(  \mathcal{C}\right)  \right\Vert _{1}\leq
4\sqrt[4]{\epsilon}.
\]
The above inequality implies that the following one holds for the expected
channel input density operators $\overline{\rho}^{\prime A^{\prime n}}$ and
$\overline{\rho}^{A^{\prime n}}$%
\begin{equation}
\left\Vert \overline{\rho}^{\prime A^{\prime n}}-\overline{\rho}^{A^{\prime
n}}\right\Vert _{1}\leq4\sqrt[4]{\epsilon},\label{eq:expected-close}%
\end{equation}
because the trace distance is convex. The following inequality holds%
\begin{equation}
\left\Vert \overline{\rho}^{A^{\prime n}}-\rho^{A^{\prime n}}\right\Vert
_{1}\leq2\epsilon+4\sqrt[4]{\epsilon}%
\end{equation}
by applying the triangle inequality to (\ref{eq:typical-close}) and
(\ref{eq:expected-close}). Therefore, the random secret-key-assisted private
code is $\rho$-like.
\end{proof}

\subsection{Associating a Random Private Code with a Classical String}

Suppose that we have an ensemble $\{p(x),\rho_{x}\}_{x\in\mathcal{X}}$ of
quantum states. The density operator $\rho_{x}$ arises as the expected density
operator of another ensemble $\left\{  p\left(  y|x\right)  ,\rho
_{x,y}\right\}  $. Let $x^{n}\equiv x_{1}\cdots x_{n}$ denote a classical
string generated by the density $p(x)$ where each symbol $x_{i}\in\mathcal{X}%
$. Then there is a density operator $\sigma_{x^{n}}$ corresponding to the
string $x^{n}$ where%
\[
\rho_{x^{n}}\equiv\bigotimes_{i=1}^{n}\rho_{x_{i}}.
\]
Suppose that we label a random private code by the string $x^{n}$ and let
$\overline{\rho}_{x^{n}}^{A^{\prime n}}$ denote its expected channel input
density operator.

\begin{definition}
A random private code is $(\rho_{x^{n}})$-like if the expected channel input
density operator $\overline{\rho}_{x^{n}}^{A^{\prime n}}$ is close to the
state $\rho_{x^{n}}$:%
\[
\left\Vert \overline{\rho}_{x^{n}}^{A^{\prime n}}-\rho_{x^{n}}\right\Vert
_{1}\leq\epsilon.
\]

\end{definition}

\begin{proposition}
\label{prop:random-grandfather} Suppose we have an ensemble as above. Consider
a quantum channel $\mathcal{N}^{A^{\prime}\rightarrow B}$ with its isometric
extension $U_{\mathcal{N}}^{A^{\prime}\rightarrow BE}$. Then there exists a
random $(\rho_{x^{n}})$-like secret-key-assisted private code for the channel
$\mathcal{N}^{A^{\prime}\rightarrow B}$ for any $\epsilon,\delta>0$, for all
sufficiently large $n$, and for any classical string $x^{n}$ in the typical
set $T_{\delta}^{X^{n}}$ \cite{CT91}. Its private communication rate is
$I(Y;B|X)-c^{\prime}\delta$, and its secret key consumption rate is
$I(Y;E|X)-c^{\prime\prime}\delta\ $for some constants $c^{\prime}%
,c^{\prime\prime}$ where the entropic quantities are with respect to the state
in (\ref{eq:maximization-state}). The state $\rho_{x}$ is the restriction of
the following state%
\[
\rho_{x}^{YA^{\prime}}=\sum_{y}p(y|x)|y\rangle\langle y|^{Y}\otimes\rho
_{x,y}^{A^{\prime}}%
\]
to the $A^{\prime}$ system.
\end{proposition}

\begin{proof}
[Proposition~\ref{prop:random-grandfather}] The proof of this theorem proceeds
exactly as the proof of Proposition~3 in Ref.~\cite{HW08GFP} and the proof of
Proposition~5 in Ref.~\cite{DS03}.
\end{proof}

\subsection{Publicly-enhanced secret-key-assisted private code}

\begin{proposition}
[HSW Coding Theorem \cite{Hol98,SW97}]\label{prop:HSW}Consider an input
ensemble $\{p(x),\rho_{x}^{A^{\prime}}\}$ that gives rise to a
classical-quantum state $\sigma^{XB}$, where%
\[
\sigma^{XB}\equiv\sum_{x\in\mathcal{X}}p(x)|x\rangle\langle x|^{X}%
\otimes\mathcal{N}^{A^{\prime}\rightarrow B}(\rho_{x}^{A^{\prime}}).
\]
Let $R=I(X;B)_{\sigma}-c^{\prime}\delta$ for any $\delta>0$ and for some
constant $c^{\prime}$. Then for all $\epsilon>0$ and for all sufficiently
large $n$, there exists a classical encoding map%
\[
h:\left[  2^{nR}\right]  \rightarrow T_{\delta}^{X^{n}},
\]
and a decoding POVM%
\[
\{\Lambda_{k}^{B^{n}}\}_{k\in\lbrack2^{nR}]},
\]
that allows Bob to decode any classical message $k\in\lbrack2^{nR}]$ with high
probability:
\[
\tr\{\tau_{k}^{B^{n}}\Lambda_{k}^{B^{n}}\}\geq1-\epsilon.
\]
The density operators $\tau_{k}^{B^{n}}$ are the channel outputs%
\begin{equation}
\tau_{k}^{B^{n}}\equiv\mathcal{N}^{A^{\prime n}\rightarrow B^{n}}(\rho
_{h(k)}^{A^{\prime n}}),\label{eq:channel-outputs}%
\end{equation}
and the channel input states $\rho_{x^{n}}^{A^{\prime n}}$ are a tensor
product of states in the ensemble:%
\[
\rho_{x^{n}}^{A^{\prime n}}\equiv{\bigotimes\limits_{i=1}^{n}\ }\rho_{x_{i}%
}^{A^{\prime}}.
\]

\end{proposition}

We are now in a position to prove the direct coding part of the
publicly-enhanced private father capacity theorem. The proof is similar to
that in Ref.~\cite{DS03,HW08GFP}.

\begin{proof}
[Direct Coding Theorem] Define the public message set $[2^{nR}]$, the
classical encoding map $h$, the channel output states $\tau_{k}^{B^{n}}$, and
the decoding POVM $\{\Lambda_{k}^{B^{n}}\}_{k\in2^{nR}}$ as in Proposition
\ref{prop:HSW}. We label each public message $k\in\lbrack2^{nR}]$ where
$R=I(X;B)-c^{\prime}\delta$.

Invoking Proposition~\ref{prop:random-grandfather}, there exists a random
$(\rho_{h(k)}^{A^{\prime n}})$-like private code $\mathcal{C}_{k}$ with
probability density $p_{\mathcal{C}_{k}}$ because each input to the channel
$\rho_{h(k)}^{A^{\prime n}}$ is a tensor product of an ensemble $\{p(x),\rho
_{x}^{A^{\prime}}\}$. The random private code $\mathcal{C}_{k}$ has
encryption-decryption pair $(f_{\mathcal{C}_{k}},g_{\mathcal{C}_{k}})$ and
encoding-decoding pair $\left(  \mathcal{E}_{\mathcal{C}_{k}},\mathcal{D}%
_{\mathcal{C}_{k}}\right)  $ for each of its realizations. We label the
combined operations simply as the pair $(\mathcal{E}_{\mathcal{C}_{k}}%
^{MS_{A}\rightarrow A^{\prime n}},\mathcal{D}_{\mathcal{C}_{k}}^{B^{n}%
S_{B}\rightarrow M})$. It transmits $n[I(Y;B|X)+c^{\prime}\delta]$ private
bits, provided Alice and Bob share at least $n[I(Y;E|X)+c^{\prime\prime}%
\delta]$ secret key bits.

Let $\mathcal{C}$ denote the \textit{random publicly-enhanced
secret-key-assisted private code} that is the collection of random private
codes $\{\mathcal{C}_{k}\}_{k\in\lbrack2^{nR}]}$. We first prove that the
expectation of the error probability for public message $k$ is small. The
expectation is with respect to the random private code $\mathcal{C}_{k}$. Let
$\tau_{\mathcal{C}_{k}}^{B^{n}}$ denote the \textit{channel output density
operator} corresponding to the private code $\mathcal{C}_{k}$:
\[
\tau_{\mathcal{C}_{k}}^{B^{n}}\equiv\text{Tr}_{S_{B}}\left\{  \mathcal{N}%
^{A^{\prime n}\rightarrow B^{n}}(\mathcal{E}_{\mathcal{C}_{k}}^{MS_{A}%
\rightarrow A^{\prime n}}(\pi^{M}\otimes\overline{\Phi}^{S_{A}S_{B}%
}))\right\}  .
\]
Let $\overline{\tau}_{k}^{B^{n}}$ denote the \textit{expected channel output
density operator} of the random father code $\mathcal{C}_{k}$:%
\[
\overline{\tau}_{k}^{B^{n}}\equiv\mathbb{E}_{\mathcal{C}_{k}}\left\{
\tau_{\mathcal{C}_{k}}^{B^{n}}\right\}  =\sum_{\mathcal{C}_{k}}p_{\mathcal{C}%
_{k}}\tau_{\mathcal{C}_{k}}^{B^{n}}.
\]
The following inequality holds%
\[
\left\Vert \overline{\rho}_{h(k)}^{A^{\prime n}}-\rho_{h(k)}^{A^{\prime n}%
}\right\Vert _{1}\leq\left\vert \mathcal{X}\right\vert \epsilon
\]
because the random private code $\mathcal{C}_{k}$ is $(\rho_{h(k)}^{A^{\prime
n}})$-like. Then the expected channel output density operator $\overline{\tau
}_{k}^{B^{n}}$ is close to the tensor product state $\tau_{k}^{B^{n}}$ in
(\ref{eq:channel-outputs}):%
\begin{equation}
\left\Vert \overline{\tau}_{k}^{B^{n}}-\tau_{k}^{B^{n}}\right\Vert _{1}%
\leq\left\vert \mathcal{X}\right\vert \epsilon,
\label{eq:channel-product-closeness}%
\end{equation}
because the trace distance is monotone under the quantum operation
$\mathcal{N}^{A^{\prime n}\rightarrow B^{n}}$. It then follows that the POVM
element $\Lambda_{k}^{B^{n}}$ has a high probability of detecting the expected
channel output density operator $\overline{\tau}_{k}^{B^{n}}$:%
\begin{align}
\tr\{\Lambda_{k}^{B^{n}}\overline{\tau}_{k}^{B^{n}}\}  &  \geq\tr\{\Lambda
_{k}^{B^{n}}\tau_{k}^{B^{n}}\}-\left\Vert \overline{\tau}_{k}^{B^{n}}-\tau
_{k}^{B^{n}}\right\Vert _{1}\nonumber\\
&  \geq1-\epsilon-\left\vert \mathcal{X}\right\vert \epsilon.
\label{eq:average-output-op-good}%
\end{align}
The first inequality follows from the following lemma that holds for any two
quantum states $\rho$ and $\sigma$ and a positive operator $\Pi$ where
$0\leq\Pi\leq I$:%
\[
\text{Tr}\left\{  \Pi\rho\right\}  \geq\text{Tr}\left\{  \Pi\sigma\right\}
-\left\Vert \rho-\sigma\right\Vert _{1}.
\]
The second inequality follows from Proposition~\ref{prop:HSW} and
(\ref{eq:channel-product-closeness}). Let $p_{e,\text{pub}}(\mathcal{C}_{k})$
denote the public message error probability for each public message $k$ of the
publicly-enhanced father code $\mathcal{C}$:%
\[
p_{e,\text{pub}}(\mathcal{C}_{k})\equiv1-\Pr\{K^{\prime}=k\ |\ K=k\}.
\]
Then by the above definition, and (\ref{eq:average-output-op-good}), it holds
that the expectation of the error probability $p_{e,\text{pub}}(\mathcal{C}%
_{k})$ for public message $k$ with respect to the random private code
$\mathcal{C}_{k}$ is low:%
\begin{align}
\mathbb{E}_{\mathcal{C}_{k}}\left\{  p_{e,\text{pub}}(\mathcal{C}%
_{k})\right\}   &  =1-\tr\{\Lambda_{k}^{B^{n}}\overline{\tau}_{k}^{B^{n}%
}\}\label{eq:classical-error}\\
&  \leq\left(  1+|\mathcal{X}|\right)  \epsilon.
\end{align}

We now show that the private error is small. Input the state $\pi^{M}%
\otimes\overline{\Phi}^{S_{A}S_{B}}$ to the encoder $\mathcal{E}%
_{\mathcal{C}_{k}}^{MS_{A}\rightarrow A^{\prime n}}$, followed by the channel
$\mathcal{N}^{A^{\prime n}\rightarrow B^{n}}$.\ The resulting state is an
extension $\Omega_{\mathcal{C}_{k}}^{S_{B}B^{n}}$ of $\tau_{\mathcal{C}_{k}%
}^{B^{n}}$:%
\[
\Omega_{\mathcal{C}_{k}}^{S_{B}B^{n}}\equiv\mathcal{N}^{A^{\prime
n}\rightarrow B^{n}}\left(  \mathcal{E}_{\mathcal{C}_{k}}^{MS_{A}\rightarrow
A^{\prime n}}(\pi^{M}\otimes\overline{\Phi}^{S_{A}S_{B}})\right)  .
\]
Let $\overline{\Omega}_{k}^{S_{B}B^{n}}$ denote the expectation of
$\Omega_{\mathcal{C}_{k}}^{S_{B}B^{n}}$ with respect to the random code
$\mathcal{C}_{k}$:%
\[
\overline{\Omega}_{k}^{S_{B}B^{n}}\equiv\mathbb{E}_{\mathcal{C}_{k}}\left\{
\Omega_{\mathcal{C}_{k}}^{S_{B}B^{n}}\right\}  .
\]
It follows that $\overline{\Omega}_{k}^{S_{B}B^{n}}$\ is an extension of
$\overline{\tau}_{k}^{B^{n}}$. The following inequality follows from
(\ref{eq:average-output-op-good}):%
\begin{equation}
\tr\{\overline{\Omega}_{k}^{S_{B}B^{n}}\Lambda_{k}^{B^{n}}\}\geq
1-(1+|\mathcal{X}|)\epsilon.
\end{equation}
The above inequality is then sufficient for us to apply a modified version of
the gentle measurement lemma\ (See Appendix~C of Ref.~\cite{HW08GFP})\ so
that\ the following inequality holds%
\begin{align}
&  \mathbb{E}_{\mathcal{C}_{k}}\left\{  \left\Vert \sqrt{\Lambda_{k}^{B^{n}}%
}\Omega_{\mathcal{C}_{k}}^{S_{B}B^{n}}\sqrt{\Lambda_{k}^{B^{n}}}%
-\Omega_{\mathcal{C}_{k}}^{S_{B}B^{n}}\right\Vert _{1}\right\}  \nonumber\\
&  \leq\sqrt{8(1+|\mathcal{X}|)\epsilon}.\label{gm}%
\end{align}
We define a decoding instrument $\mathcal{D}_{\mathcal{C}}^{B^{n}%
S_{B}\rightarrow KM}$ for the random publicly-enhanced private father code
$\mathcal{C}$\ as follows \cite{Yard05a,HW08GFP}:%
\begin{align*}
&  \mathcal{D}_{\mathcal{C}}^{B^{n}S_{B}\rightarrow KM}\left(  \rho
^{B^{n}S_{B}}\right)  \\
&  \equiv\sum_{k}\mathcal{D}_{\mathcal{C}_{k}}^{B^{n}S_{B}\rightarrow
M}\left(  \sqrt{\Lambda_{k}^{B^{n}}}\rho^{B^{n}S_{B}}\sqrt{\Lambda_{k}^{B^{n}%
}}\right)  \otimes\left\vert k\right\rangle \left\langle k\right\vert ^{K},
\end{align*}
where $\mathcal{D}_{\mathcal{C}_{k}}^{B^{n}S_{B}\rightarrow M}$ is the decoder
for the private father code $\mathcal{C}_{k}$ and each map $\mathcal{D}%
_{\mathcal{C}_{k}}^{B^{n}S_{B}\rightarrow M}(\sqrt{\Lambda_{k}^{B^{n}}}%
\rho^{B^{n}S_{B}}\sqrt{\Lambda_{k}^{B^{n}}})$ is trace-reducing. The induced
quantum operation corresponding to this instrument is as follows:%
\[
\mathcal{D}_{\mathcal{C}}^{B^{n}S_{B}\rightarrow M}\left(  \rho\right)
=\text{Tr}_{K}\left\{  \mathcal{D}_{\mathcal{C}}^{B^{n}S_{B}\rightarrow
KM}\left(  \rho\right)  \right\}  .
\]
Monotonicity of the trace distance gives an inequality for the trace-reducing
maps of the quantum decoding instrument:%
\begin{align}
&  \mathbb{E}_{\mathcal{C}_{k}}\left\{  \left\Vert
\begin{array}
[c]{c}%
\mathcal{D}_{\mathcal{C}_{k}}^{B^{n}S_{B}\rightarrow M}\left(  \sqrt
{\Lambda_{k}^{B^{n}}}\Omega_{\mathcal{C}_{k}}^{S_{B}B^{n}}\sqrt{\Lambda
_{k}^{B^{n}}}\right)  -\\
\mathcal{D}_{\mathcal{C}_{k}}^{B^{n}S_{B}\rightarrow M}\left(  \Omega
_{\mathcal{C}_{k}}^{S_{B}B^{n}}\right)
\end{array}
\right\Vert _{1}\right\}  \nonumber\\
&  \leq\sqrt{8(1+|\mathcal{X}|)\epsilon}.\label{eq:epsilon-sqrt-map}%
\end{align}
The following inequality also holds%
\begin{align}
&  \mathbb{E}_{\mathcal{C}_{k}}\left\{  \left\Vert
\begin{array}
[c]{c}%
\mathcal{D}_{\mathcal{C}}^{B^{n}S_{B}\rightarrow M}\left(  \Omega
_{\mathcal{C}_{k}}^{S_{B}B^{n}}\right)  -\\
\mathcal{D}_{\mathcal{C}_{k}}^{B^{n}S_{B}\rightarrow M}\left(  \sqrt
{\Lambda_{k}^{B^{n}}}\Omega_{\mathcal{C}_{k}}^{S_{B}B^{n}}\sqrt{\Lambda
_{k}^{B^{n}}}\right)
\end{array}
\right\Vert _{1}\right\}  \nonumber\\
&  \leq\mathbb{E}_{\mathcal{C}_{k}}\left\{  \sum_{k^{\prime}\neq k}\left\Vert
\mathcal{D}_{\mathcal{C}_{k^{\prime}}}^{B^{n}S_{B}\rightarrow M}\left(
\sqrt{\Lambda_{k^{\prime}}^{B^{n}}}\Omega_{\mathcal{C}_{k}}^{S_{B}B^{n}}%
\sqrt{\Lambda_{k^{\prime}}^{B^{n}}}\right)  \right\Vert _{1}\right\}
\nonumber\\
&  =\mathbb{E}_{\mathcal{C}_{k}}\left\{  \sum_{k^{\prime}\neq k}\left\Vert
\sqrt{\Lambda_{k^{\prime}}^{B^{n}}}\Omega_{\mathcal{C}_{k}}^{S_{B}B^{n}}%
\sqrt{\Lambda_{k^{\prime}}^{B^{n}}}\right\Vert _{1}\right\}  \nonumber\\
&  =\mathbb{E}_{\mathcal{C}_{k}}\left\{  \sum_{k^{\prime}\neq k}%
\text{Tr}\left\{  \Lambda_{k^{\prime}}^{B^{n}}\Omega_{\mathcal{C}_{k}}%
^{S_{B}B^{n}}\right\}  \right\}  \nonumber\\
&  =1-\text{Tr}\left\{  \Lambda_{k}^{B^{n}}\overline{\Omega}_{k}^{S_{B}B^{n}%
}\right\}  \nonumber\\
&  \leq(1+|\mathcal{X}|)\epsilon.\label{eq:epsilon-diff-maps}%
\end{align}
The first inequality follows by definitions and the triangle inequality. The
first equality follows because the trace distance is invariant under isometry.
The second equality follows because the operator $\Lambda_{k}^{B^{n}}%
\Omega_{\mathcal{C}_{k}}^{S_{B}B^{n}}$ is positive. The third equality follows
from some algebra, and the second inequality follows from
(\ref{eq:average-output-op-good}). The private communication for all public
messages $k$ and codes $\mathcal{C}_{k}$ is good
\[
\left\Vert \mathcal{D}_{\mathcal{C}_{k}}^{B^{n}S_{B}\rightarrow M}\left(
\Omega_{\mathcal{C}_{k}}^{S_{B}B^{n}}\right)  -\pi^{M}\right\Vert _{1}%
\leq\epsilon,
\]
because each code $\mathcal{C}_{k}$\ in the random private father code is good
for private communication. It then follows that%
\begin{equation}
\mathbb{E}_{\mathcal{C}_{k}}\left\{  \left\Vert \mathcal{D}_{\mathcal{C}_{k}%
}^{B^{n}S_{B}\rightarrow M}\left(  \Omega_{\mathcal{C}_{k}}^{S_{B}B^{n}%
}\right)  -\pi^{M}\right\Vert _{1}\right\}  \leq\epsilon
.\label{eq:good-q-comm}%
\end{equation}
Application of the triangle inequality to (\ref{eq:good-q-comm}),
(\ref{eq:epsilon-diff-maps}), and (\ref{eq:epsilon-sqrt-map}) gives the
following bound on the expected private error probability:%
\begin{equation}
\mathbb{E}_{\mathcal{C}_{k}}\left\{  p_{e,\text{priv}}\left(  \mathcal{C}%
_{k}\right)  \right\}  \leq\epsilon^{\prime}\label{eq:private-error}%
\end{equation}
where%
\[
\epsilon^{\prime}\equiv(1+|\mathcal{X}|)\epsilon+\sqrt{8(1+|\mathcal{X}%
|)\epsilon}+2\sqrt{\epsilon},
\]
and where we define the private error $p_{e,\text{priv}}\left(  \mathcal{C}%
_{k}\right)  $ of the code $\mathcal{C}_{k}$ as follows:%
\[
p_{e,\text{priv}}\left(  \mathcal{C}_{k}\right)  \equiv\left\Vert
\mathcal{D}_{\mathcal{C}}^{S_{B}\rightarrow M}\left(  \Omega_{\mathcal{C}_{k}%
}^{S_{B}B^{n}}\right)  -\pi^{M}\right\Vert _{1}.
\]

The above random publicly-enhanced secret-key-assisted private code relies on
Alice and Bob having access to a source of common randomness. We now show that
they can eliminate the need for common randomness and select a good
publicly-enhanced secret-key-assisted private code $\mathcal{C}$ that has a
low public error $p_{e,\text{pub}}(\mathcal{C}_{k})$ and low private error
$p_{e,\text{priv}}(\mathcal{C}_{k})$ for all public messages in a large subset
of $[2^{nR}]$. By the bounds in (\ref{eq:classical-error}) and
(\ref{eq:private-error}), the following bound holds for the expectation of the
averaged summed error probabilities:
\[
\mathbb{E}_{\mathcal{C}_{k}}\left\{  \frac{1}{2^{nR}}\sum_{k}p_{e,\text{pub}%
}(\mathcal{C}_{k})+p_{e,\text{priv}}(\mathcal{C}_{k})\right\}  \leq
\epsilon^{\prime}+(1+|\mathcal{X}|)\epsilon.
\]
If the above bound holds for the expectation over all random codes, it follows
that there exists a particular publicly-enhanced private father code
$\mathcal{C}=\left\{  \mathcal{C}_{k}\right\}  _{k\in\left[  2^{nR}\right]  }$
with the following bound on its averaged summed error probabilities:%
\[
\frac{1}{2^{nR}}\sum_{k}p_{e,\text{pub}}(\mathcal{C}_{k})+p_{e,\text{priv}%
}(\mathcal{C}_{k})\leq\epsilon^{\prime}+(1+|\mathcal{X}|)\epsilon.
\]
We fix the code $\mathcal{C}$ and expurgate the worst half of the private
father codes---those private father codes with public messages $k$ that have
the highest value of $p_{e,\text{pub}}(\mathcal{C}_{k})+p_{e,\text{priv}%
}(\mathcal{C}_{k})$. This derandomization and expurgation yields a
publicly-enhanced private father code that has each public error
$p_{e,\text{pub}}(\mathcal{C}_{k})$ and each private error $p_{e,\text{priv}%
}(\mathcal{C}_{k})$ upper bounded by $2\left(  \epsilon^{\prime}%
+(1+|\mathcal{X}|)\epsilon\right)  $ for the remaining public messages $k$.
This expurgation decreases the public rate by a negligible factor of $\frac
{1}{n}$.
\end{proof}

\section{Child Protocols}

\label{sec:children}Two simple protocols for the public-private setting are
\textit{secret key distribution} and the \textit{one-time pad} \cite{V26,S49}.
Secret key distribution is a protocol where Alice creates the state
$\overline{\Phi}^{AA^{\prime}}$ locally and sends the system $A^{\prime}$
through a noiseless private channel. The protocol creates a secret key and
corresponds to the following resource inequality:%
\[
\left[  c\rightarrow c\right]  _{\text{priv}}\geq\left[  cc\right]
_{\text{priv}}.
\]
The one-time pad protocol exploits a secret key and a noiseless public channel
to create a noiseless private channel. It admits the following resource
inequality:%
\[
\left[  c\rightarrow c\right]  _{\text{pub}}+\left[  cc\right]  _{\text{priv}%
}\geq\left[  c\rightarrow c\right]  _{\text{priv}}.
\]

We now consider some protocols that are child protocols of the
publicly-enhanced private father protocol. Consider the resource inequality in
(\ref{eq:PEPFP-resource-ineq}). We can combine the protocol with secret key
distribution, and we recover the protocol suggested in Section~4 of
Ref.~\cite{DS03}:%
\begin{align*}
&  \left\langle \mathcal{N}\right\rangle +I\left(  Y;E|X\right)  _{\sigma
}\left[  cc\right]  _{\text{priv}}\\
&  \geq I\left(  Y;B|X\right)  _{\sigma}\left[  c\rightarrow c\right]
_{\text{priv}}+I\left(  X;B\right)  _{\sigma}\left[  c\rightarrow c\right]
_{\text{pub}}.\\
&  \geq\left(  I\left(  Y;B|X\right)  _{\sigma}-I\left(  Y;E|X\right)
_{\sigma}\right)  \left[  c\rightarrow c\right]  _{\text{priv}}+\\
&  I\left(  Y;E|X\right)  _{\sigma}\left[  c\rightarrow c\right]
_{\text{priv}}+I\left(  X;B\right)  _{\sigma}\left[  c\rightarrow c\right]
_{\text{pub}}\\
&  \geq\left(  I\left(  Y;B|X\right)  _{\sigma}-I\left(  Y;E|X\right)
_{\sigma}\right)  \left[  c\rightarrow c\right]  _{\text{priv}}+\\
&  I\left(  Y;E|X\right)  _{\sigma}\left[  cc\right]  _{\text{priv}}+I\left(
X;B\right)  _{\sigma}\left[  c\rightarrow c\right]  _{\text{pub}}%
\end{align*}
By cancellation of the secret key term, we are left with the following
resource inequality:%
\begin{align*}
\left\langle \mathcal{N}\right\rangle +o\left[  cc\right]  _{\text{priv}}  &
\geq\left(  I\left(  Y;B|X\right)  _{\sigma}-I\left(  Y;E|X\right)  _{\sigma
}\right)  \left[  c\rightarrow c\right]  _{\text{priv}}\\
&  +I\left(  X;B\right)  _{\sigma}\left[  c\rightarrow c\right]  _{\text{pub}%
},
\end{align*}
where $o\left[  cc\right]  _{\text{priv}}$ represents a sublinear amount of
secret key consumption.

We can combine the publicly-enhanced private father protocol with the one-time
pad:%
\begin{align}
&  \left\langle \mathcal{N}\right\rangle +I\left(  Y;E|X\right)  _{\sigma
}\left[  cc\right]  _{\text{priv}}+I\left(  X;B\right)  _{\sigma}\left[
cc\right]  _{\text{priv}}\nonumber\\
&  \geq I\left(  Y;B|X\right)  _{\sigma}\left[  c\rightarrow c\right]
_{\text{priv}}+I\left(  X;B\right)  _{\sigma}\left[  c\rightarrow c\right]
_{\text{pub}}\nonumber\\
&  +I\left(  X;B\right)  _{\sigma}\left[  cc\right]  _{\text{priv}}\\
&  \geq I\left(  Y;B|X\right)  _{\sigma}\left[  c\rightarrow c\right]
_{\text{priv}}+I\left(  X;B\right)  _{\sigma}\left[  c\rightarrow c\right]
_{\text{priv}}\nonumber\\
&  =I\left(  XY;B\right)  _{\sigma}\left[  c\rightarrow c\right]
_{\text{priv}}\label{eq:combine-OTP}%
\end{align}
This protocol is one for secret-key-assisted transmission of private
information. It is not an efficient protocol because the optimal
secret-key-assisted protocol \cite{HLB08SKP}\ implements the following
resource inequality:%
\[
\left\langle \mathcal{N}\right\rangle +I\left(  XY;E\right)  _{\sigma}\left[
cc\right]  _{\text{priv}}\geq I\left(  XY;B\right)  _{\sigma}\left[
c\rightarrow c\right]  _{\text{priv}}%
\]
For a channel with non-zero private capacity so that $I\left(  X;B\right)
_{\sigma}-I\left(  X;E\right)  _{\sigma}>0$, the protocol in
(\ref{eq:combine-OTP}) is not efficient because it uses more secret key than
necessary. This inefficiency is similar to the inefficiency that we found for
combining the classically-enhanced father protocol with teleportation (See
Section~VII of Ref.~\cite{HW08GFP}). It is not surprising that this
inefficiency occurs because the publicly-enhanced private father protocol is
the public-private analog of the classically-enhanced father protocol and the
one-time pad protocol is the public-private analog of the teleportation
protocol \cite{CP02}.

\section{Conclusion}

We have introduced an optimal protocol, the publicly-enhanced private father
protocol, that exploits a secret key and a large number of independent uses of
a noisy quantum to transmit public and private information. Several protocols
in the literature are now special cases of this protocol.

A few open questions remain. It remains to determine the capacity regions of a
multiple-access quantum channel \cite{YHD05ieee,itit2008hsieh} and a broadcast
channel \cite{YHD2006}\ for transmitting public and private information while
consuming a secret key. One might also consider the five-dimensional region
corresponding to the scenario where Alice and Bob consume secret key,
entanglement, and a noisy quantum channel to produce quantum communication,
public classical communication, and private classical communication. This
scenario might give more insight into the privacy/coherence correspondence. It
remains open to determine the full triple trade-off for the use of a quantum
channel in connection with public communication, private communication, and
secret. We have made initial progress on this problem by exploiting techniques
developed in Ref.~\cite{HW09T3}. Before completing this work, we need to
determine a publicly-assisted private mother protocol, the analog of the
classically-assisted mother protocol in Refs.~\cite{DHW05RI,HW09T3}. This
protocol should then allow us to determine the full triple trade-off for both
the dynamic setting and the static setting.

\begin{acknowledgments}
The authors thank Igor Devetak for a private discussion regarding the issue in
Section~\ref{sec:relative-resource-priv}\ with the protocol for private
communication. MMW\ acknowledges partial support from an internal research and
development grant SAIC-1669 of Science Applications International Corporation.
\end{acknowledgments}

\bibliographystyle{unsrt}
\bibliography{Ref}

\end{document}